\documentclass[aps,prapplied,amsmath,amssymb,amsfonts,lengthcheck,twocolumn,superscriptaddress]{revtex4-2}
\usepackage{graphicx}
\usepackage{subfigure}
\usepackage{amsthm}
\usepackage{verbatim}
\usepackage{dcolumn}
\usepackage{bm}
\usepackage{epsf}
\usepackage{color}
\usepackage[colorlinks=true,citecolor=blue,linkcolor=blue,urlcolor=blue]{hyperref}%
\usepackage{xcolor}
\usepackage{dsfont}
\usepackage{tikz}
\usepackage{todonotes}
\usepackage{multirow}
\usepackage{xfrac}
\usepackage{makecell}

\newcommand{\bla}{bla\\bla\\bla\\bla\\bla}

\makeatletter
\newcommand{\currentfontsize}{\f@size pt}
\makeatother

\makeatletter
\newcommand\footnoteref[1]{\protected@xdef\@thefnmark{\ref{#1}}\@footnotemark}
\makeatother

\begin{document}

\title{On the Baltimore Light RailLink into the quantum future}

\author{Krzysztof Domino}
\thanks{These authors contributed equally.}
\email{kdomino@iitis.pl}

\affiliation{Institute of Theoretical and Applied Informatics, Polish Academy of Sciences, Ba{\l}tycka 5, Gliwice, 44-100, Poland}

\author{Emery Doucet}
\thanks{These authors contributed equally.}
\email{emery.doucet@umbc.edu}
\affiliation{Department of Physics, University of Maryland, Baltimore County, Baltimore, MD 21250, USA}
\affiliation{Quantum Science Institute, University of Maryland, Baltimore County, Baltimore, MD 21250, USA}

\author{Reece Robertson}
\affiliation{Department of Computer Science and Electrical Engineering, University of Maryland, Baltimore County, Baltimore, MD 21250, USA}
\affiliation{Department of Physics, University of Maryland, Baltimore County, Baltimore, MD 21250, USA}
\affiliation{Quantum Science Institute, University of Maryland, Baltimore County, Baltimore, MD 21250, USA}

\author{Bartłomiej Gardas}
\affiliation{Institute of Theoretical and Applied Informatics, Polish Academy of Sciences, Ba{\l}tycka 5, Gliwice, 44-100, Poland}

\author{Sebastian Deffner}
\homepage{https://quthermo.umbc.edu/} 
\affiliation{Department of Physics, University of Maryland, Baltimore County, Baltimore, MD 21250, USA}
\affiliation{Quantum Science Institute, University of Maryland, Baltimore County, Baltimore, MD 21250, USA}
\affiliation{National Quantum Laboratory, College Park, MD 20740, USA}

\begin{abstract}
	In the current era of noisy intermediate-scale quantum (NISQ) technology, quantum devices present new avenues for addressing complex, real-world challenges including potentially NP-hard optimization problems. Acknowledging the fact that quantum methods underperform classical solvers, the primary goal of our research is to demonstrate how to leverage quantum noise as a computational resource for optimization. This work aims to showcase how the inherent noise in NISQ devices can be leveraged to solve such real-world problems effectively. Utilizing a D-Wave quantum annealer and IonQ's gate-based NISQ computers, we generate and analyze solutions for managing train traffic under stochastic disturbances. Our case study focuses on the Baltimore Light RailLink, which embodies the characteristics of both tramway and railway networks. We explore the feasibility of using NISQ technology to model the stochastic nature of disruptions in these transportation systems. Our research marks the inaugural application of both quantum computing paradigms to tramway and railway rescheduling, highlighting the potential of quantum noise as a beneficial resource in complex optimization scenarios.
\end{abstract}


\maketitle

\section{Introduction}

It is an important goal to realize scalable and fault-tolerant quantum computers \cite{pirnay2023superpolynomial,divincenzo2000physical,Sanders2017}. However, the technological challenges are immense and currently available devices are characterized and governed by noise. These noisy intermediate-scale quantum (NISQ) devices \cite{preskill2018quantum} are already capable of practical computations, yet any application will have to embrace numerical imperfections and shortcomings. Thus, the obvious and  important question arises whether the inevitably noisy characteristics can be exploited as a resource.

The applicability of quantum computing has already been demonstrated in optimization problems for industrial and operations research tasks \cite{katzgraber2024searching}.  A particular area of interest is railway transportation, where, for instance, a few hours of operation of a single metro line with about $13$ stations and $150$ trains was solved on a current classical computer in approximately $5$ hours. In practice, re-scheduling and dispatching decisions must be made in seconds \cite{bevsinovic2020resilience} and hence new technologies such as a quantum approach are desirable \cite{zhou2022joint,ge2022robustness,bevsinovic2020resilience}. 

Only rather recently, employing NISQ for railway scheduling has been demonstrated~\cite{domino2023quantum}. In this work, re-scheduling was expressed as a quadratic unconstrained binary optimization (QUBO) problem, which was then solved using a D-Wave quantum annealer -- for the conceptual extension of the model to a higher order unconstrained binary optimization (HOBO) approach see~\cite{domino2022quadratic}. 
Importantly, while ideally the output of a quantum annealer is the exact optimal solution (corresponding to the ground state of the Ising model used to encode the QUBO), higher-lying excited states are often returned instead. In terms of the original scheduling problem, these higher-lying states may correspond to rail traffic with additional disturbances. Referring to railways see Ref.~\cite{xu2023high} (high-speed trains timetable optimization on coherent Ising machine quantum simulator), Ref.~\cite{grozea2021optimising} (rolling stock planning on the quantum annealer for German Railways Network), and Refs.~\cite{grange2024design, grange2022formulation} (finding the rail transportation plan on the IBM 32-qubit QASM simulator using QUBO/HOBO and quantum approximate optimization algorithm - QAOA - encoding) for a complementary approaches. However, these studies did not tie the noise of NISQ device with the stochastic behavior of a particular public transportation system. 

In the present work, we demonstrate the utility of NISQ devices  (and systematically compare annealers and gate-based computers) for the problem of railway scheduling by leveraging the inherent noise present in all such devices as a tool to model stochastic behavior in real-world problems, an idea that will prove especially relevant to the development of hybrid quantum/classical solution methods -- see Supplementary Information for pseudo code. Such hybrid methods can tackle significantly larger problems than either the classical or NISQ approach by themselves~\cite{callison2022hybrid}.

Technically, our approach follows the ideas presented in Ref.~\cite{dixit2023quantum} but with a focus on the practical application of railway re-scheduling. We focus on rail systems with some stochastic features as can be found in tramways, which partially share infrastructure with road traffic.  Some research has been performed on this problem \cite{zeng2012collaboration} and on stochastic optimization of tramway scheduling in general, e.g., in the Hong Kong tram network \cite{lai2018real}, however literature on this topic is sparse \cite{ge2022robustness} and generally with a different focus such as infrastructure repair scheduling \cite{kiefer2018scheduling} or vehicle and crew rescheduling \cite{malucelli2019delay}. In our previous research, non-optimal but practically sound solutions of railway scheduling~\cite{koniorczyk2023application} and analogous Automated Guided Vehicles scheduling~\cite {smierzchalski2024hybrid} problems have been obtained from D-Wave's proprietary hybrid solvers. Unfortunately, these solvers are closed black boxes and it isn't easy to assess the actual effort done by the QPU during optimization. Given such experience, in this work we concentrate on pure quantum computing with trackable performance.

In our work, we focus on a specific real-world transportation system: the Baltimore Light RailLink in Baltimore, Maryland \cite{BaltimoreLightRailLink}
The Baltimore Light RailLink system is an important factor in the economic development in the urban area surrounding the network \cite{barry2012economic}, and current data on its performance \cite{BaltimoreLightRailLinkPerfImprovement}
indicates that there is a margin for improvement.
These properties of the Baltimore Light RailLink infrastructure are in no way unique and are found in other tramway infrastructures (for example that in Karlsruhe, Germany \cite{kraskiewicz2015tramwaj}), therefore the conclusions we draw in this work are general.

\begin{figure}
	\centering
	\includegraphics[width = 0.5\textwidth]{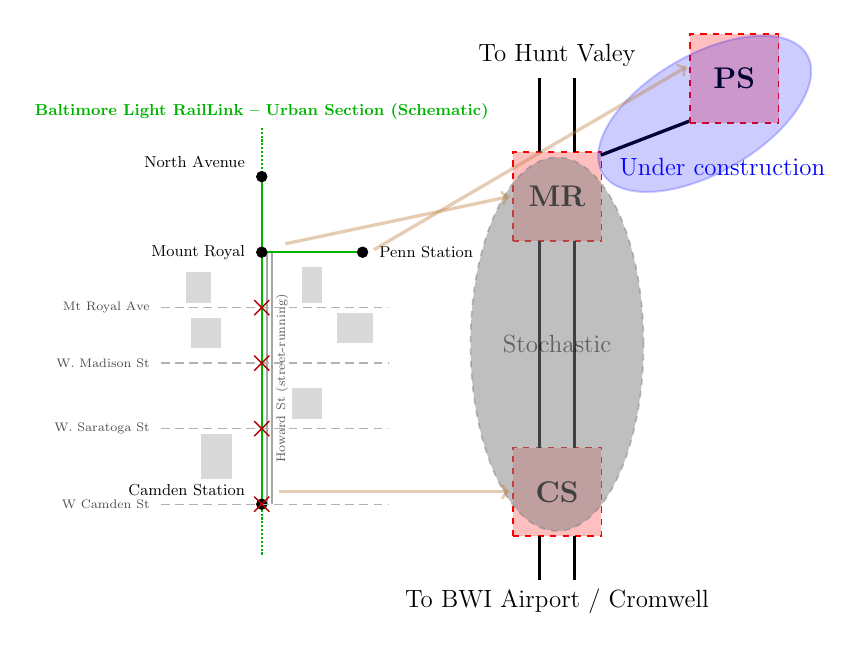}
	\caption{Map based on \cite{BLRLMapORM}
		(left) and schematic representation 
		(right) of the Baltimore Light RailLink system we consider. The stochastic zone with shared infrastructure runs between Camden Station (CS) and Mount Royal (MR). There is then a deterministic connection with dedicated infrastructure connecting Mount Royal to Penn Station (PS) - Figure generated by \emph{LaTeX tikz (pdfTeX 3.141592653-2.6-1.40.25)}.
	}\label{fig::analyzed_network}
\end{figure}

In Fig.~\ref{fig::analyzed_network} we depict the portion of the Baltimore Light RailLink network we consider in this work on a map and schematically. 
Between Camden Station and Mount Royal the trains are subject to essentially random variations in transit time due to road traffic, hence we denote the segment of the network bounded by these two stations as the ``stochastic zone''. Outside of this region, for example north to Mount Royal, the trains travel on dedicated infrastructure and travel times are deterministic. 
In this work, we analyze dispatching situations for this portion of the network in terms of \emph{decision stations} \cite{koniorczyk2023application}, with a focus on the passing time of trains through the stochastic zone.

Our approach can serve as one component of a multilevel combinatorial optimization approach combining the strengths of existing classical solvers with quantum architectures \cite{ushijima2021multilevel} for the purposes of railway scheduling, especially when the underlying network exhibits some form of stochasticity. These multilevel algorithms allow the solution of problems an order of magnitude larger than current quantum hardware can handle directly, and we expect that a hybrid algorithm based on our techniques can be applicable to large and practically-relevant railway problems, for example of size comparable with these in Ref.~\cite{zhou2022joint}.

This article is organized as follows: In Sec.~\ref{sec::model} we present the model of the Baltimore Light RailLink and solution methods for both the quantum annealing and gate-based quantum computing paradigms. In Sec.~\ref{sec::comp_results} we present results from experiments with each of these approaches, using machines from D-Wave and IonQ, alongside simulations. We discuss these results in light of real-life data measured from railway traffic, before presenting our conclusions in Sec.~\ref{sec::conclusions}. 
Additional details on encoding constraints in QUBO problems, on a proposed hybrid algorithm that can be applied to larger railway scheduling problems, on hardware details and embeddings or circuits used for our experiments with a D-Wave quantum annealer and IonQ's gate-based trapped ion computer, and on some tests using IBM's superconducting gate-based computers are available as supplementary material.

\section{Mathematical models}\label{sec::model}

The basic idea behind our research is to apply the NISQ device to model the trains' traffic in the stochastic zone in Fig.~\ref{fig::analyzed_network}, as quantum computers are NISQ devices~\cite{preskill2018quantum} and so their output is inherently noisy and stochastic. 
For encoding this problem in a form amenable to quantum computing, we follow the outline of~\cite{domino2023quantum}. 
We first encode the scheduling problem as an Integer Linear Programming (ILP) problem, which is then transformed into a Quadratic Unconstrained Binary Optimization (QUBO) problem which serves as the input for various optimization methods developed for quantum computers.

\subsection{Integer Linear Programming approach}\label{sec::linear_model}

The state-of-the-art approach \cite{bevsinovic2020resilience} to railway scheduling -- and re-scheduling in particular -- is to encode the problem as an ILP problem and solve it using high-quality solvers such as CPLEX. 
Stochastic components are often handled by solving many ILP problems with various parameters~\cite{lai2018real}, or with simulations, see e.g.~\cite{kroon2008stochastic}. 
In this section, we present the mapping of our train scheduling problem to an ILP problem.
To do so it is necessary to specify the variables which are used to encode a solution timetable, the constraints imposed on those variables, and finally the objective used to evaluate solution timetables for optimality. 
Throughout, we will concentrate primarily on re-scheduling problems where the input timetable requires modification, for example, due to it being infeasible as a result of disturbances causing constraint violations.

\subsubsection{Variables}
To begin, we assign numeric identifiers to the trains we wish to schedule and to the stations they will visit. 
We will use the subscript $j$ to select a particular train, and $\#J$ to denote the total number of trains. 
Similarly, the subscript $s$ will select a specific station and $\#S$ will indicate the total number of stations.

We introduce a set of non-negative integer variables
\begin{equation}
	t_{s,j}^{\text{in}} \in \mathbb{Z}^+,
	\label{eq::t_vars_def}
\end{equation}
which represent the time train $j$ enters station $s$. 
As is typical for railway scheduling, we measure times with a resolution of one minute \cite{bevsinovic2020resilience}. 
Note that if $s$ is the first station in the train's $j$ route, then $t_{s,j}^{\text{in}}$ is the time the train is ready to start its route.
After the minimal stop at the station, it can proceed.

We denote the arrival time of train $j$ at station $s$ as given by the not-disturbed timetable by $\tau^{\rm in}_{s,j}$.
The earliest time at which the train could arrive is denoted by $l_{s,j} \geq \tau^{\rm in}_{s,j}$. 
This is just the non-disturbed timetable arrival time if the train has not been delayed, or if the input timetable is infeasible the lowest possible arrival time given no rail traffic.
If we assume that there is a maximal allowed additional (called also secondary \cite{DarianoPDPbb}) delay given by $d_{\rm max}$, then we may also define a latest possible arrival time $u_{s,j} = l_{s,j} + d_{\rm max}$ and so
\begin{equation}
	l_{s,j} \leq t_{s,j}^{in} \leq u_{s,j} 
	\quad\textrm{equivalently}\quad t
	_{s,j}^{in} \in R_{s,j}.
	\label{eq::t_range}
\end{equation}
Importantly, the maximal delay parameter $d_{\max}$ determines the problem size by constraining the range of time variables. 
We have at most $ \# S \# J$ such time variables, as we do not consider recirculation~\cite{pinedo2012scheduling}.

For the purposes of our model, we assume that the stay time of the train at the station is constant and equal to $\delta^{\text{station}}$.
We then define the time at which train $j$ leaves station $s$,
\begin{equation}
	t_{s,j}^{\text{out}} = t_{s,j}^{\text{in}} + \delta^{\text{station}}.\label{eq::min_stay_equal}
\end{equation}

Alongside the variables we have just defined which describe the times at which trains arrive at and depart from stations, we introduce binary variables which specify the precedence of trains. 
Specifically, define 
\begin{equation}
	y_{j,j',s} \in \{0, 1\}
	\label{eq::y_vars_def}
\end{equation}
so that $y_{j,j',s} = 1$ if train $j$ enters station $s$ before train $j'$ (by considering the reversed case, we see that $y_{j,j',s} = 1 - y_{j',j,s}$). 

Since we have imposed a maximal delay time of $d_{\rm max}$, then following the discussion in \cite{koniorczyk2023application} we expect that each train has a dependency on (i.e., is potentially in conflict with) a roughly constant number of trains, regardless of how large the problem is. Hence, the number of $y$ variables is proportional to $\# J $ and $ \# S$.

\subsubsection{Constraints}\label{ILP::constraints}

We impose three types of constraints which must be satisfied for a solution to represent a valid schedule. 
Additionally, as all trains move with similar average speeds and have the same priority we expect that they will not meet and overtake at or between stations. This condition is not encoded explicitly but will be checked for any solution.

The first of constraint we consider is the \textbf{minimal headway} constraint, which is taken to be deterministic.
If two trains are heading in the same direction, they are required to maintain a minimal headway time $\delta^{\text{headway}}$ between them. 
Explicitly, let $H_{s}$ be the set of all pairs of trains which have a minimal headway dependency entering $s$. 
Then:
\begin{equation}
	\forall_s \forall_{(j,j') \in H_{s}} \begin{cases} 
		t_{s,j'}^{\text{in}} \geq t_{s,j}^{\text{in}} + \delta^{\text{headway}} &\text{ if } y_{j,j',s} = 1 \\ 
		t_{s,j}^{\text{in} } \geq t_{s,j'}^{\text{in}} + \delta^{\text{headway}} &\text{ if } y_{j,j',s} = 0.
	\end{cases}
	\label{eq::min_headway}
\end{equation}
From this, we conclude that we have one minimal headway constraint per $y$ variable. 
The average size of $H_{s}$ over the set of stations depends on the $d_{\max}$ parameter and is expected to be proportional to the number of trains.

The second constraint is the \textbf{rolling stock circulation} constraint, which is also taken to be deterministic.
If two trains are heading in opposite directions, they may use the same rolling stock so be dependent. In such a case, if the train $j$ terminates at $s$, then after a preparation time $\delta^{\text{preparation}}$ the subsequent train $j'$ is ready to start its journey backwards. 
Let $RS_s$ be the set of all such pairs of trains, which is timetable-dependent. 
Then:
\begin{equation}
	\forall_{(j,j') \in RS_s} t^{\text{in}}_{s,j'} \geq t^{\text{in}}_{s,j} + \delta^{\text{preparation}} + \delta^{\text{station}}.\label{eq::min_circ}
\end{equation}
There are far fewer rolling stock circulation constraints than there are minimal headway constraints.

The final constraint we consider is a constraint on the \textbf{minimal passing time}, and it is here where the underlying stochasticity of the model appears. 
In the deterministic scheduling case, the minimal passing time between stations is given by the train-independent constant $\delta^{\text{pass}}_{s,s'}$. Then the minimal passing time constraint implies:
\begin{equation}
	\forall_j \forall_{(s,s') \in SP_j} t^{\text{in}}_{s',j} \geq t^{\text{out}}_{s,j} + \delta^{\text{pass}}_{s,s'},
	\label{eq::min_pass}
\end{equation}
where $SP_j$ is a set of all pairs of subsequent stations in the route of $j$. The average size (over trains) of $SP_j$, namely $|SP_j|$, is expected to be proportional to the total number of stations in the model. 

As argued when introducing our model shown in Fig.~\ref{fig::analyzed_network}, the effect of road traffic on the shared infrastructure will be to add some randomness to the passing time through the stochastic zone.
To model this, we modify Eq.~\eqref{eq::min_pass} to:
\begin{equation}
	\forall_j \forall_{(s,s') \in SP_j} t^{\text{in}}_{s',j} \geq t^{\text{out}}_{s,j} + \delta^{\text{pass}}_{s,s'} + w.
	\label{eq::min_pass_stochastic}
\end{equation}
where $w \in \mathcal{W}$ is the particular realization of a non-negative stochastic factor representing the additional unpredictable delay. 
To gather statistics of the problem, such an approach would require solving many ILP problems that employ the constraints of Eq.~\eqref{eq::min_pass_stochastic} with a range of actual values of $w$, as there are $|\mathcal{W}|$ such possibilities for each inequality. In the worst case, we expect to need to solve approximately $\#J \# S |\mathcal{W}|$ ILP problems, which could be a very large number. 
In this work, we demonstrate that a similar outcome can be achieved from a single set of runs of a quantum device, which can be performed in seconds.

All together, from Eq.~\eqref{eq::min_pass} and Eq.~\eqref{eq::min_headway} and the surrounding discussion we expect the number of constraints to be linear in $\# J$ and $\# S$.

\subsubsection{Objective}
The objective function we seek to minimize when constructing a timetable is the sum of delays at some selected stations listed in $S^*$,
\begin{equation}
	f = \sum_j \sum_{s \in S^*} \frac{t^{\text{in}}_{s, j} - \tau^{\text{in}}_{s, j}}{d_{\max}}.
	\label{eq::objective_linear}
\end{equation}
We assume that the deterministic approach with minimal passing time constraints given by Eq.~\eqref{eq::min_pass} yields the optimal solution.  
As $w$ in Eq.~\eqref{eq::min_pass_stochastic} is non-negative, the stochastic approach may result in feasible solutions with higher values for the objective $f$. 
This is also true of the output of a minimization using a NISQ device. 
The remainder of this paper is dedicated to examining this observation.

\subsection{QUBO approach for quantum computing}\label{sec::QUBO_model}

The ILP problem formulated in the previous section defines our scheduling problem, however it is not in a form which is easily solvable with NISQ devices. 
For that, it is necessary to move to a QUBO representation (for a detailed discussion of the QUBO formulation, see \cite{glover2019quantum}).
Following \cite{domino2023quantum}, we define the optimization problem in terms of the binary decision variables
\begin{equation}
	x_{s,j,t} = \begin{cases} 1 \text{ if train } j \text{ enters station } s \text{ at time } t \\ 0 \text{ otherwise} \end{cases}.
	\label{eq::bin_v}
\end{equation}
where $t$ is in the interval $R_{s,j}$ defined in Eq.~\eqref{eq::t_range}.
The number of these decision variables $N$ is at most $\# J \# S (d_{\max} + 1)$, as some trains may serve not all stations.

The binary decision variables in Eq.~\eqref{eq::bin_v} can be gathered into a vector $\vec{x} \in \{0, 1\}^{N}$, where the elements of $\vec{x}$ are labeled $x_i$, where each $i$ is the flattened version of the corresponding multi-index $s,j,t$ introduced in Eq.~\eqref{eq::bin_v}.
Then, the (re-)scheduling problem becomes equivalent to finding the vector $\vec{x}$ which minimizes the quadratic form
\begin{equation}
	E(\vec{x})=\vec{x} Q \vec{x}^T \in \mathbb{R}
	,
	\label{eq::Q}
\end{equation}
to determine the minimum objective value $E_0 = \min_{\vec{x}} E(\vec{x})$ and optimal solution $\vec{x}_0 = \textrm{argmin} E(\vec{x})$. $Q$ is the matrix encoding the constraints and objective function, which can be taken to be symmetric. 
The details on encoding the minimal passing time [cf. Eq.~\eqref{eq::min_pass}] minimal headway [cf. Eq.~\eqref{eq::min_headway}] and rolling stock circulation [cf. Eq.~\eqref{eq::min_circ}] constraints as discussed in Section~\ref{ILP::constraints} in terms of a quadratic form on the binary decision variables are presented in 
the Supplementary Information. 
Ultimately, we expect both the number of qubits ($N$) and the number of non-zero entries in the $Q$ matrix to be proportional to $\# J \# S$. Hence, for constant $d_{\text{max}}$, the mean degree of the problem graph is expected to be independent on $\# J $ and $ \# S$. 

These constraints are all quadratic and have the form
\begin{equation}
	\sum_{i,i'} x_{i} x_{i'} = 0.
	\label{eq::qubo_pair0}
\end{equation}
Transforming from the constrained optimization problem of the original scheduling problem and subsequent ILP formulation to an unconstrained problem which can be encoded as a QUBO is performed by making the hard constraints soft with an associated penalty for constraint violations,
\begin{equation}
	\sum_{i< i'} p_{\text{pair}} (x_{i} x_{i'} + x_{i'} x_{i}).\label{eq::qubo_pair}
\end{equation}
Once expressed in this form, it is straightforward to input the constraints into the $Q$ matrix of Eq.~\eqref{eq::Q}.
Following~\cite{domino2023quantum}, we chose one value $p_{\text{pair}}$ for all quadratic terms in Eq.~\eqref{eq::qubo_pair} associated with violations of the constraints, but the generalization is straightforward. The number of these constraints is expected to be linear in $\# S$ $\# J$, as discussed in Section~\ref{ILP::constraints} and 
the Supplementary Information. 

We have another set of constraints, the so-called one-hot constraints \cite{venturelli2015quantum}. 
These constraints ensure that each train leaves each station once and only once:
\begin{equation}
	\forall_{s,j \in (S,J)}  \sum_{ t \in R_{j,s} } x_{s,j,t} = 1,
	\label{eq::bin_sum}
\end{equation}
wherein the set of pairs of trains and stations is denoted by $(S,J)$ and where $t$ runs over the range $R_{j,s}$ introduced in Eq.~\eqref{eq::t_range}.
The number of these sums is limited by $\# S \# J$, as at most each train passes each station.
This can be written in unconstrained form with a penalty, 
\begin{equation}
	\sum_{s,j \in (S,J)}  p_{\text{sum}} \left( \sum_{t \in R_{j,s} } x_{s,j,t} - 1\right)^2,
	\label{eq::qubo_sum}
\end{equation}
which is then transformed using the identity $x_{s,j,t}^2 = x_{s,j,t}$ yielding finally,
\begin{equation}
	\begin{split}
		\sum_{s,j \in (S,J)} p_{\text{sum}} \Bigg(\sum_{\substack{t, t' \in R_{j,s} \\ t \neq t'}} x_{s,j,t} x_{s,j,t'} - \sum_{t \in R_{j,s} } x_{s,j,t} ^2 \Bigg).
		\label{eq::bin_sum_uncontrained}
	\end{split}
\end{equation}
We expect approximately $(d_{\max} + 1)^2 \# J \# S$ such terms.

The objective is linear and is directly derived from Eq.~\eqref{eq::objective_linear}:
\begin{equation}
	f(\vec{x}) = \sum_j \sum_{s \in S^*} \left( \sum_{ t \in R_{j,s}} \frac{t - \tau^{\text{in}}_{s,j}}{d_{\max}} x_{s, j, t} \right).
\end{equation}

The solution to the QUBO of Eq.~\eqref{eq::Q} will correspond to the optimal solution to the scheduling problem. 
We expect that nearly optimal solutions whose objective values are not much larger will correspond to solutions which are consistent with the constraints but which include increased passing time between stations. 
Solutions further from optimal with larger values of the objective function will then typically correspond to solutions which fail to satisfy the pair and sum constraints of the forms presented in Eq.~\eqref{eq::qubo_pair} and Eq.~\eqref{eq::qubo_sum}, respectively.
The interplay between these two situations is controlled by the penalty parameters in the QUBO, $p_{\text{sum}}$ and $p_{\text{pair}}$.

Identifying good values $p_{\text{sum}}$ and $p_{\text{pair}}$ for the penalty parameters is a complicated problem, and recent years have seen the development of dedicated algorithms for this purpose \cite{alessandroni2023alleviating}.
These algorithms have, however, complexity that must be included for in estimates of the computational time of the final problem (e.g., the general procedure of finding the proper quantum encoding/circuits is NP-hard itself~\cite{bittel2021training}). 
Furthermore, algorithmic approaches are typically based on deriving a lower bound for the penalty coefficients necessary to enforce the various constraints by splitting feasible and non-feasible parts of the spectrum.
Then, one expects the penalty from the violation of constraint to be larger than any possible changes of the objective. 
In practice, better results are often recorded for penalty coefficients smaller than the above-mentioned lower limits \cite{liu2022leveraging}. 
In this case, the spectra corresponding to the feasible and non-feasible solution spaces overlap.
We expect that this yields some positive impact on the solution process, e.g. by smoothing local minima in the quantum evolution process. 

To understand this effect, the studies presented in this work were all repeated twice. In one case which we call the \emph{overlapping spectrum} case, we choose the penalty values such that one broken constraint corresponds to $4\times$ the worst objective at a single train and station, $p_{\text{sum}} =4$ and $p_{\text{pair}} = 2$. The other case we call the \emph{split spectrum} case, where the penalties are made $10\times$ larger: $p_{\text{sum}} =40$ and $p_{\text{pair}} = 20$.

Recall that if we explicitly require that trains do not meet and overtake at or between
stations (see Section~\ref{ILP::constraints}), a HOBO encoding such as in Ref.\cite{domino2022quadratic} would be required. However, such an approach would substantially complicate the problem for the annealer, leading to rather limited practical benefit in our case. For example, in Ref.\cite{domino2022quadratic} HOBO problems corresponding to at most $3$ trains and $2$ stations were handled successfully by pure quantum annealing on the D-Wave machine. Nevertheless, we acknowledge, that the HOBO encoding can be more competitive when applying the gate-based quantum computing via the QAOA approach, which opens a particular avenue for further research.

\subsection{Ising Model for Quantum annealing}

The final step necessary to express the scheduling problem in a form amicable to solution on a NISQ device is to transform from QUBO to Ising representation. 
This reformulation is straightforwardly implemented by introducing the spin variables $s_i \in \{-1, 1\}$ and taking $s_i = 2 x_i - 1$, where $x_i$ are the variables in the original QUBO representation. In this way, the QUBO formulation in Eq.~\eqref{eq::Q} is transformed into the Ising Hamiltonian:
\begin{equation}
	H(\vec{s}) = \sum_{(i,i')} J_{i,i'} s_i s_{i'} + \sum_{i} s_i h_i.
	\label{eq::ising_hamiltonian}
\end{equation}
Here $J_{i,i'}$ and $h_i$ are couplings and constant terms resulting from particular $Q$ in QUBO in Eq~\eqref{eq::Q}. 
Minimization of the latter over spin configurations is consistent with the minimization of the corresponding 
Eq.~\eqref{eq::Q} with binary variables.

Quantum annealers such as those manufactured by D-Wave implement the quantum version of the Ising model, where spin variables become Pauli matrices acting on spin's subspace.
The Ising representation is the native input of the quantum annealer, where the original problem in Eq.~\eqref{eq::ising_hamiltonian} turns into the search for the lowest-energy (or low-energy in practice) eigenstate of the following Hermitian Hamiltonian:
\begin{equation}
	H = \sum_{(i,i')} J_{i,i'} \sigma^z_i \sigma^z_{i'} + \sum_{i} \sigma^z_i h_i.
	\label{eq::ising_hamiltonian_q}
\end{equation}
The spin variable $s_i = \pm 1$ in Eq.~\eqref{eq::ising_hamiltonian} is associated with $\sigma^z_i$ via $\sigma_i^z |s_i\rangle = s_i|s_i\rangle$, where $|s_i\rangle$ form a local computational basis.

The basic idea of quantum annealing relies on the celebrated adiabatic theorem~\cite{avron1999adiabatic}.
One prepares the initial state of the quantum system in the ground state of the initial Hamiltonian $H_0 =  \sum_{i} \sigma^x_i h_i$, then slowly evolves the Hamiltonian to $H_p$ (the evolution time being called the annealing time $T$). 
If the evolution is sufficiently slow, then the final state will ideally be the ground state or at least a low-lying excited state of the final Hamiltonian, Eq.~\eqref{eq::ising_hamiltonian_q} in this case, and hence a low-energy state of the original Ising model of Eq.~\eqref{eq::ising_hamiltonian}.
The performance of this process is dependent on a variety of factors, e.g. on the eigenvalues of the Hamiltonian of Eq.~\eqref{eq::ising_hamiltonian_q} and their spacing. 
These relationships are complex, and it is not obvious what will be observed when analyzing two QUBOs with different penalty coefficients and hence different spectra.

As a final note, the Ising Hamiltonians corresponding to our scheduling problems are defined on a dense graph.
Real quantum annealers do not have all-to-all connectivity, and so mapping our problems onto the native coupling graph of an actual quantum annealer~\cite{dattani2019pegasus} requires the use of a standard procedure called ''minor embedding''. 
This embedding leads the overhead in the number of quantum bits and limits the size of the problem that can be handled by any given quantum annealer. 
As the size of the problem graph is linear in $\# J$ and $\# S$ in terms of number of logical qubits and number of connections, we expect the number of physical qubits to be also proportional to $\# J$ and $\# S$. 
Hence, we do not expect an explosion in the required size of quantum processor necessary for large problems of practical relevance.

\subsection{Model for Gate Quantum Computing}

Besides quantum annealing, the most common approaches to solving combinatorial optimization problems on NISQ-era quantum computers use the gate-based paradigm of quantum computing with variational algorithms. These algorithms make use of standard optimization algorithms running on a classical computer, with a quantum computer being used to evaluate the objective function which is to be minimized.
There are various such algorithms \cite{cerezo2021variational, blekos2023review}, which employ various variational ansatzes and are tuned for different purposes.

In this work, we use the quantum approximate optimization algorithm (QAOA) \cite{farhi2014quantum}.
Given an $N$-qubit diagonal Hamiltonian $H_C$ representing a cost function to be minimized (such as the Ising Hamiltonian encoding the scheduling problem given by Eq.~\eqref{eq::ising_hamiltonian_q}) and a mixing Hamiltonian (typically $H_M = \sum_{i} \sigma^x_i$), the algorithm proceeds by using a quantum computer to generate the ansatz state
\begin{equation}
	|\vec{\beta}, \vec{\gamma}\rangle = e^{-i\beta_p H_M}e^{-i\gamma_p H_C}\dots e^{-i\beta_1 H_M}e^{-i\gamma_1 H_C}|+\rangle^{\otimes N} 
\end{equation}
and compute the expectation value of the cost Hamiltonian $\langle \vec{\beta}, \vec{\gamma}|H_C|\vec{\beta},\vec{\gamma}\rangle$ where $\vec{\beta}$ and $\vec{\gamma}$ are $p$-dimensional vectors of coefficients characterizing each of the $p$ layers of the ansatz.
Once the $2p$ parameters that minimize this expectation value are identified, the corresponding ansatz state provides a good approximation of the ground state of the cost Hamiltonian, and so the prepared ansatz state in the computational basis yields an approximate solution to the initial combinatorial optimization problem.
To find the optimal parameters, the computation of the expectation value using a quantum computer is treated as a black box function to be optimized with some classical optimization method, for example, the COBYLA method \cite{powell1994direct} which has been shown to work well on these types of problems \cite{fernandezpendas2022study}.

When comparing the process of solving a QUBO using QAOA against using a quantum annealer such as D-Wave, it is clear that QAOA is, in some sense, a more expensive process.
One optimization using the algorithm—essentially equivalent to a single shot on a quantum annealer—requires a classical optimization to be run. This requires that a quantum computer be used to run a series of circuits with different parameter values to extract the objective function. Each of these requires many independent shots to sample the prepared ansatz to compute the objective function reliably.

It is nonetheless interesting to compare the performance of a general-purpose gate-based quantum computer against a dedicated quantum annealer on combinatorial optimization problems. 
While QAOA is inspired by adiabatic computation with a quantum annealer, it is not clear if the two models should be expected to exhibit similar behaviour as various problem parameters are adjusted (e.g., the problem size or the choice of penalty parameters). We also expect that the effect of noise present in current NISQ-era implementations may not affect the optimization result from a gate-based computer implementing QAOA in the same way as a quantum annealer.

As a final note, our decision to employ QAOA as our algorithm of choice for solving our problem instances using gate-based quantum hardware is motivated primarily by simplicity, efficacy, and ease of comparison to the results of quantum annealing. 
One of the major limitations of this approach is the limited qubit count available with presently-available gate-based quantum computers.
This limitation can be mitigated in several ways, for example by considering hybrid quantum-classical optimizers that only solve subproblems on the quantum hardware (as we discuss in the Supplementary Information), or by considering clever encoding schemes that allow the solution of large problems with few qubits \cite{tan2021qubit, tene2023variational, sciorilli2025towards}.
Such approaches are especially important on current NISQ hardware, as smaller circuits on fewer qubits promise higher accuracy and better optimization performance.
An performance advantage due to encoding may also be possible if instead of distilling the scheduling problem to a QUBO, it is encoded as a higher-order binary optimization problem (HOBO). QAOA applied to higher-order problems can require fewer qubits, at the cost of increased circuit depth \cite{glos20space, stein2023evidence}. 

Embracing the gate-based paradigm completely, there also exist approaches capable of solving scheduling problems without initially reducing the problem to a QUBO, for instance through quantum versions of the branch-and-bound algorithm \cite{montanaro2020quantum, chakrabarti2022universal}.
These algorithms pose a challenge for current devices, but in the future may be a promising avenue towards solving large-scale problems with gate-based quantum hardware.

\section{Computational results}\label{sec::comp_results}

To test our approach, we concentrate primarily on the stochastic zone of the Baltimore Light RailLink identified in the introduction and depicted in Fig.~\ref{fig::analyzed_network}.
This stochastic zone is the central part of the Baltimore Light RailLink between station Camden Station (CS) and Mount Royal (MR) station where railway traffic is shared with road traffic, and is connected to Penn Station (PS) with a section of track, currently under reconstruction.
We model these three stations as \emph{decision stations}~\cite{koniorczyk2023application}, the stations at which scheduling decisions are made. 

From a real Baltimore Light RailLink timetable \cite{BaltimoreLightRailLinkSchedule}
we can derive a set of model parameters. For the whole network, these are the minimum headway time $\delta^{\text{headway}} = 2$, the minimum preparation time $\delta^{\text{preparation}} = 3$, and minimum time in station $\delta^{\text{station}} = 1$. For the passage between stations MR and PS, the minimum passing times in the two directions are $\delta^{\text{pass}}_{\text{MR, PS}} = 14$ and $\delta^{\text{pass}}_{\text{PS, MR}} = 2$.

To model train traffic on this segment, we start with the real Baltimore Light RailLink timetable without any disturbances. Note that in our model we also include several trains whose route either starts from or ends at station PS, though these trains do not run at the current time as station PS is closed due for reconstruction \cite{BaltimoreLightRailLinkConstruction}.
This allows us to model rather dense railway traffic, and so we may limit the maximum secondary delay $d_{\rm max}$ which ensures the QUBO representations are not too large.

To ensure that our analysis is as thorough as possible, we generate various scheduling problem instances which incorporate from a number of trains $\# J \in \{1,2,4,6,8,10,11,12\}$ and which have maximal delay parameters $d_{\max}$ of either 2 or 6 minutes.
We divide the scheduling problems we consider into two classes:
\begin{itemize}
	\item \textbf{Non-disturbed}, where the initial condition (input timetable) is feasible, as there are no delays of trains. In this case, the optimal solution (the solution to the deterministic scheduling problem) has an objective value of zero. Feasible solutions with non-zero objective values are expected to correspond to random disturbances in the stochastic zone.
	\item \textbf{Disturbed}, where delays have been injected into the input timetable. Here, the input problem can be infeasible even without stochastic disturbances as the delay may lead to constraint violations without adjustments to the schedule. Again, the optimal solution solves the deterministic version of the problem, and feasible solutions with slightly higher objective values are expected to correspond to random disturbances in the stochastic zone (bear in mind, that for one train instance, we do not have a disturbed case). 
\end{itemize}

In total, we consider $30$ distinct scheduling problems. When translated into the QUBO formulation, the smallest problem we consider (with a single train serving $2$ stations and a maximum delay of $2$ minutes, where the delay is counted from zero) requires $6$ variables to encode. Mapped to a quantum Ising model for quantum annealing, this corresponds to $6$ logical qubits. The largest problem (with $12$ trains, each serving $2$ or $3$ stations, and a maximum delay of $6$ minutes) requires $196$ variables, which is comparable in size to the largest railway scheduling problems solved previously on a D-Wave machine in \cite{domino2023quantum}. Note that since the D-Wave machine does not support arbitrary connectivity between its qubits, there is some overhead in embedding the QUBOs corresponding to our scheduling problems. For a deeper discussion, see the Supplementary Information. Fig.~\ref{fig::instances}a shows the relationship between the number of logical qubits required for each QUBO we consider and the number of physical qubits necessary after embedding, which can be seen to be approximately linear. Also shown in Fig.~\ref{fig::instances}b is an extrapolation of this scaling to a large enough number of logical qubits to handle problematic real-world railway scheduling problems \cite{zhou2022joint}, which are around $60\times$ larger than the largest problem we consider in this work (in terms of $\# J \# S$, e.g. including around 13 stations and 150 trains).

For each of these $30$ problems, solutions can be obtained with classical ILP solvers or through the solution of the corresponding QUBO problem on NISQ devices. In the latter case, the results will be stochastic due to the pervasive noise in current quantum computing platforms. Through the remainder of this section, we present details of the solution of these problems on a D-Wave quantum annealer and on a trapped-ion quantum computer by IonQ. 

\begin{figure}  
	\centering
	\includegraphics[width = 1.0\columnwidth]{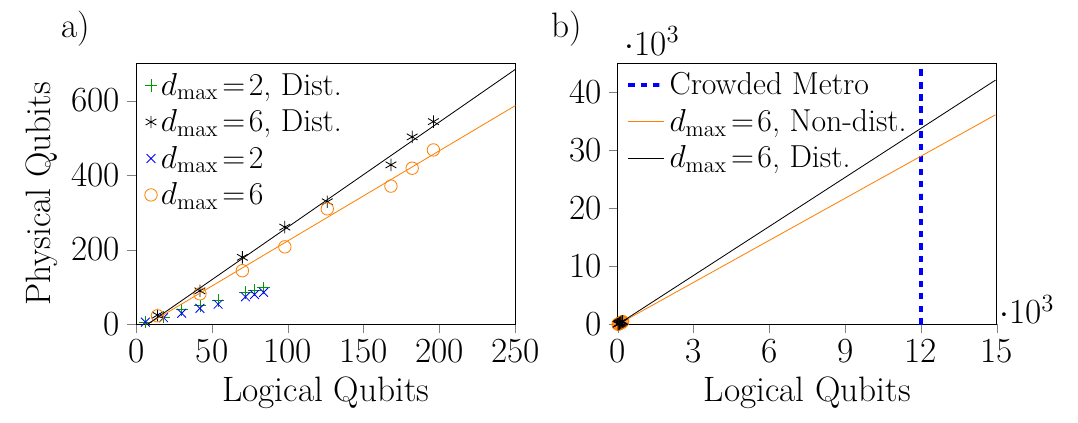}
	\caption{
		Relationship between the number of logical qubits (variables) necessary to encode the scheduling problems considered in this work as QUBOs and the number of physical qubits required after embedding into the native coupling graph of the D-Wave annealer, for disturbed and non-disturbed instances with two choices for $d_{\rm max}$. The trend for the problems we considered (a) is approximately linear, which can be extrapolated (b) to roughly estimate the number of qubits required to fit a crowded metro scheduling problem \cite{zhou2022joint}.
	}\label{fig::instances}
\end{figure}

\subsection{Quantum annealer}

For the exploration of quantum annealing in this work, we have employed the D-Wave \texttt{Advantage\_system6.3} machine with 5614 qubits and a configurable annealing time $T$ (for hardware specifics, see 
the Supplementary Information). 
For each scheduling problem, we executed $25000$ runs to build a large distribution of solutions. We did this for each of the $30$ scheduling problems we defined, each run twice with annealing times of $10{\mu s}$ and $1000{\mu s}$ to study the effects on performance. The D-Wave embedding strategy uses chains of physical qubits to represent individual logical qubits. Here, we have used the default setting of \emph{dwave.system}, the \emph{EmbeddingComposite} Python library for all calculations on the D-Wave. The actual percentage of the number of solutions with no broken chains is greater than $50 \%$ for all calculations. We acknowledge, however, that properly setting chain strengths can have a positive impact on the annealer's performance \cite{king2015performance}, and assessing this problem is the sound direction of further investigation.

\begin{figure}
	\centering
	\includegraphics[width = 1.0\columnwidth]{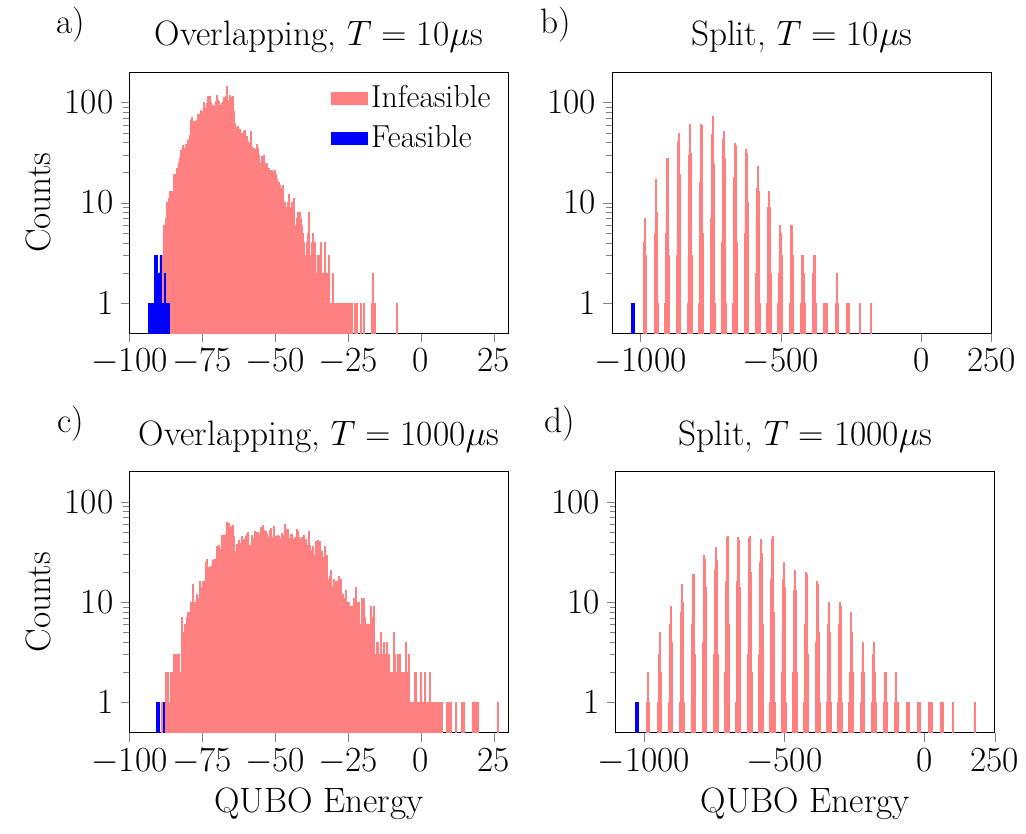}
	\caption{Histograms of result energies measured with the D-Wave annealer showing the impact of the annealing time and penalty values on the QUBOs' energy spectrum. Results are shown for the disturbed instance of $11$ trains with $d_{\max} = 6$ requiring $182$ logical and $503$ physical qubits, with annealing times of $T=10\mu{\rm s}$ (a,b) and $T=1000\mu{\rm s}$ (c,d) and penalty values $p_{\rm pair}=2, p_{\rm sum}=4$ (a,c) and $p_{\rm pair}=20, p_{\rm sum}=40$ (b,d), which yield overlapping and split spectra respectively. Each histogram shows the results of 25000 shots. Comparing the left and right panels makes the splitting of the spectrum due to the use of larger penalty values very clear. Comparing the top and bottom panels shows that in this case, a shorter annealing time typically yields better results.
	}
	\label{fig::11trains_spectra}
\end{figure}

An example of the result distributions we obtained is presented in Fig.~\ref{fig::11trains_spectra}, which shows the distribution of solution energies produced for the 11 train scheduling problem broken down by annealing time and choice of penalty parameter values, with solutions categorized as feasible and infeasible (for the 12 train problems, certain parameter settings yielded no feasible solutions.
The best results are obtained with a shorter annealing time and when the penalty values are chosen to give an overlapping spectrum as seen in Fig.~\ref{fig::11trains_spectra}a. The preference for a shorter annealing time is presumably a consequence of noise, but the improved performance when the spectrum is overlapping is a more non-trivial observation.
From it, we conclude that for larger problems, when the annealing time is short it may be that penalty values which allow the feasible and non-feasible regions of the spectrum to overlap provide some computational advantage.
A plausible intuitive explanation is given by considering that when the spectrum is split, if during the annealing process the system enters an excited state it may be locked in the excited and non-feasible region due to the spectral gap between the two regions. 

\begin{figure}
	\centering
	\includegraphics[width = 1.0\columnwidth]{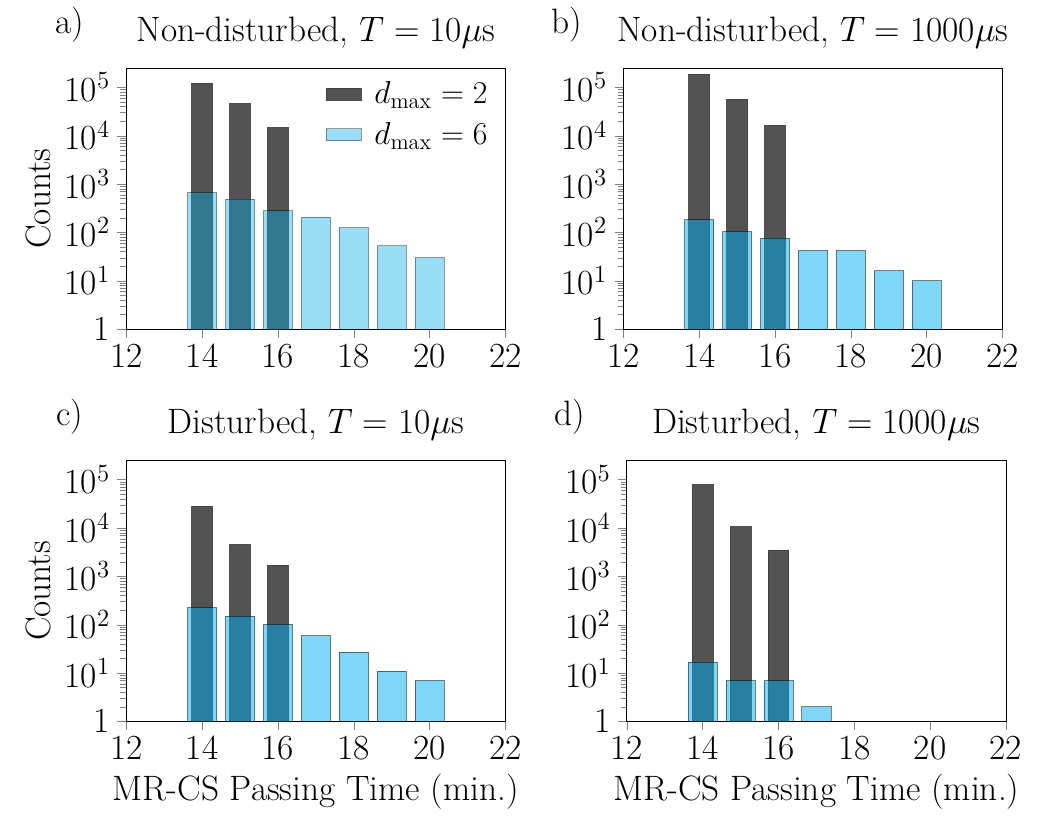}  
	\caption{
		Histograms showing the impact of the maximal secondary delay $d_{\max}$ and the annealing time $T$ on passing times through the stochastic zone computed from feasible solutions generated by the D-Wave annealer for large instances with $11$ trains ($182$ variables at $d_{\max} = 6$), both disturbed (a,b) and non-disturbed (c,d). Annealing times $T$ of $10\mu{\rm s}$ (a,c) and $1000\mu{\rm s}$ (b,d) are shown. In all cases, we use the penalty values $p_{\text{sum}} =4$ and $p_{\text{pair}} = 2$ corresponding to an overlapping spectrum.  Recall that the ILP versions of the problems were solved using CPLEX, returning a single optimal result for each instance, yielding the $14$ minutes of MR-CS passing time; the CPLEX computational times of ILP problems were in the range of $0.0016$ - $0.07$ seconds.}
	\label{fig::qubo_larger_dmax6}
\end{figure}

To leverage these results practically for the problem of rail traffic scheduling, we concentrate on the statistics of passing times through the stochastic zone from feasible solutions. For the various 11 train scheduling problems, these statistics are presented in Fig.~\ref{fig::qubo_larger_dmax6}. In a later section, these distributions will be compared to the distribution of passing times measured from the actual Baltimore Light RailLink network. 
More generally, the findings presented in Fig.~\ref{fig::qubo_larger_dmax6} qualitatively reproduce features of typical tramway delay frequency distributions due to disturbances as presented in Figs. 3, 4, 7, and 8 of \cite{ullrich2016regular}. Our results are also similar to the right tail of histograms of typical passing times of road-based public transportation as presented in Fig. 1a of \cite{buchel2020review}.

As a final note, we observe that the shape of the result distributions is sensitive to the maximal delay parameter $d_{\rm max}$. It is not just that the distribution widens, but the overall contour appears to develop additional features. This alongside the variations in the distributions due to different annealing times and penalty values indicates that it should be possible to exert a certain level of control over the shape of the passing time distribution by adjusting these parameters.

\subsection{Gate-based computers}

For the gate-based approach, we solved our scheduling problems with QAOA running on the trapped ion hardware and noisy simulator provided by IonQ.
IonQ's ``Aria-1'' device provides 25 qubits with all-to-all connectivity, meaning that there is no embedding overhead when solving any given QUBO unlike with D-Wave's quantum annealers.
We solved four non-disturbed scheduling problems with variable counts of 6, 10, 14, and 18 on the Aria-1 device and one disturbed scheduling problem with 18 variables. These problems required the execution of quantum circuits operating on 6 to 18 qubits. For details on these circuits and on the hardware properties, see 
the Supplementary Information. 
Taking the largest 18 variable QUBO, the corresponding QAOA circuit with a single-layer ansatz contained 54 single-qubit gates and 30 two-qubit gates. In terms of circuit complexity, the most challenging QUBO to solve is actually the 14 variable instance, which has a more complicated set of constraints leading to the single-layer ansatz compiling to a circuit with 42 single-qubit gates and 63 two-qubit gates.

On the Aria-1 device, the average gate error rates 
can vary considerably over time, with single-qubit gate errors generally averaging between $0.03\%$ and $0.06\%$ and with two-qubit gate errors generally averaging between $2.1\%$ and $8.6\%$. Further details on the characterization of the device are given in 
the Supplementary Information. 
To understand the impact of gate errors on the device performance for our scheduling problems, we solved each QUBO problem using QAOA with both a one- and two-layer ansatz.
In principle, the more complicated two-layer variational ansatz should allow the prepared state to better approximate the true ground state of the problem Hamiltonian and therefore should lead to better optimization performance. In practice, however, this theoretical improvement is offset by the increased circuit depth and so increased susceptibility to gate errors and decoherence. 
Here, for the largest 18 variable QUBO the two-layer QAOA circuit required 90 single- and 60 two-qubit gates.
This is approximately double the number of the simpler circuit for the single-layer ansatz.

Due to constraints on hardware availability, we solved each of the four non-disturbed scheduling problems four times with QAOA on the Aria-1 hardware, with the two ansatze and two choices of penalty parameters. 
To better understand the results, we also solved each scheduling problem (both disturbed and non-disturbed) with QAOA run on the noisy simulator $50$ times for each problem size and choice of parameters.
Since we observe reasonable agreement between the results from the simulator and the device, we can use the simulator results as a proxy to understand the expected distribution of results with the gate-based approach. 

\begin{figure}
	\centering
	\includegraphics[width = 1.0\columnwidth]{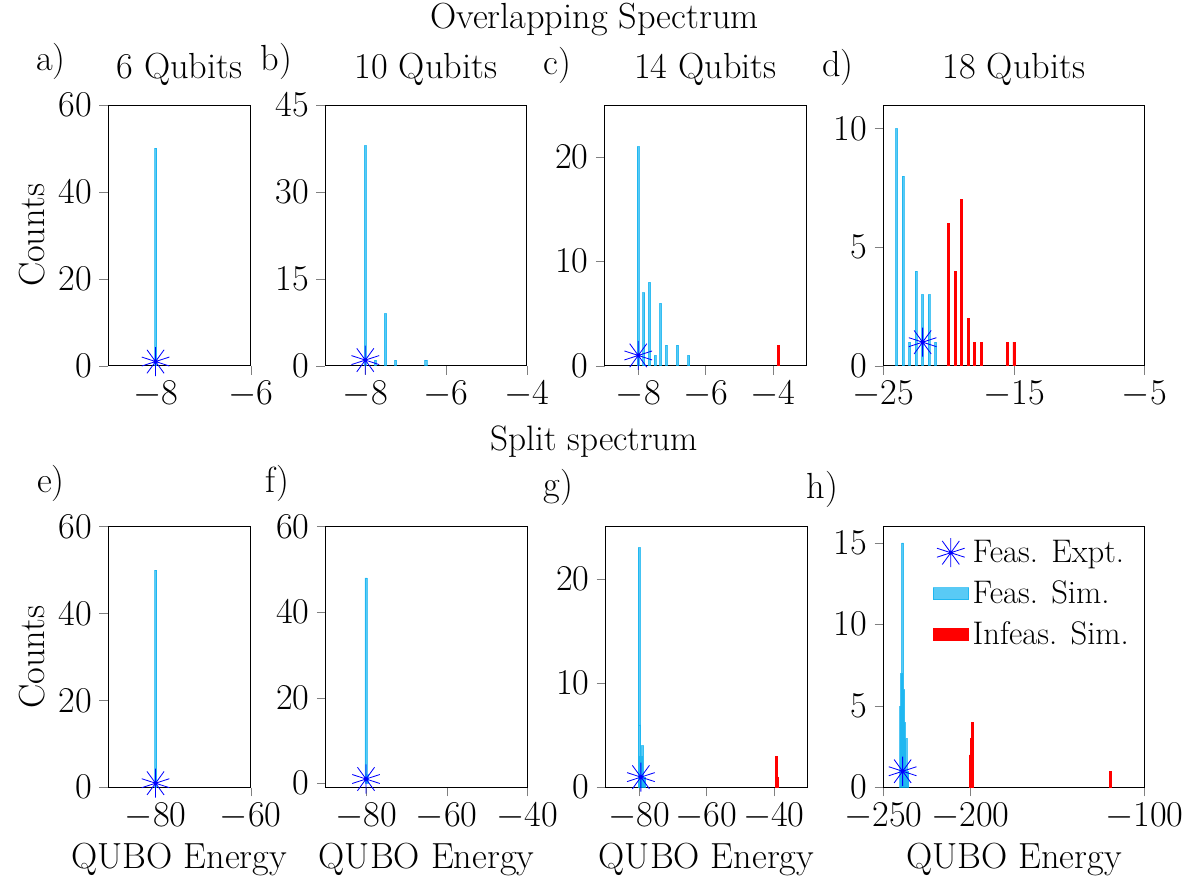}
	\caption{Histograms of the result energy measured using QAOA with a single-layer ansatz to solve the QUBOs corresponding to small non-disturbed scheduling problems on IonQ's Aria-1 machine. The histograms show the results from noisy simulations, run 50 times per problem. Each QUBO was solved once using the real hardware, the resulting energies of these trials are shown with blue stars. From these histograms, we observe better results when the penalty parameters are such that the spectrum is split (bottom).}\label{fig::Aria_real}
\end{figure}

The full set results from these experiments and simulations with a single-layer ansatz are presented in Fig.~\ref{fig::Aria_real}. Based on the fact that in each case the single experiment performed with the real hardware produced a result which was either optimal or close to optimal and which corresponded to an obvious peak in the histogram of results of the noisy simulations, we conclude that the noisy simulator does a reasonable job of capturing the behavior of the hardware and can therefore be used to make qualitative statements and predictions about its performance for train scheduling. Interestingly, the results we obtain from the noisy simulations show that the Aria-1 device performs better when the QUBO spectrum is split. This is opposite to the behavior of the D-Wave machine on the hardest instance solved as shown in Fig.~\ref{fig::11trains_spectra}.

\begin{figure}
	\centering
	\includegraphics[width = 1.0\columnwidth]{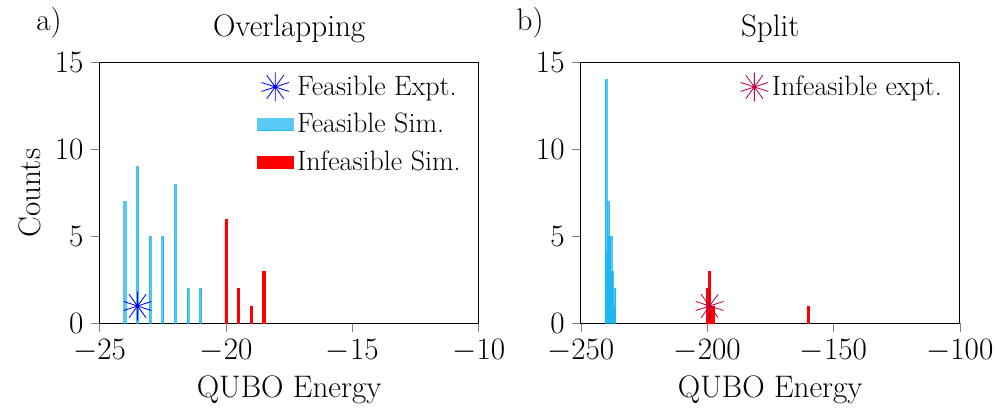}
	\caption{Histograms of the result energy measured when solving the largest 18 variable QUBOs from a non-disturbed scheduling problem using QAOA with a two-layer ansatz on IonQ's Aria-1. As in Fig.~\ref{fig::Aria_real}, the histograms represent the results of 50 noisy simulations and the blue/red stars show the results of a single trial using the actual Aria-1 device. As before, we observe better results when the spectrum is split (right).}
	\label{fig::Aria_real_2l}
\end{figure}

Fig.~\ref{fig::Aria_real_2l} shows the results of the experiments and simulations solving the largest $18$ variable problem with a two-layer ansatz running on the Aria-1 device. The results are somewhat mixed, and we can not conclude which approach is better in this case.

\begin{figure}
	\centering
	\includegraphics[width = 1.0\columnwidth]{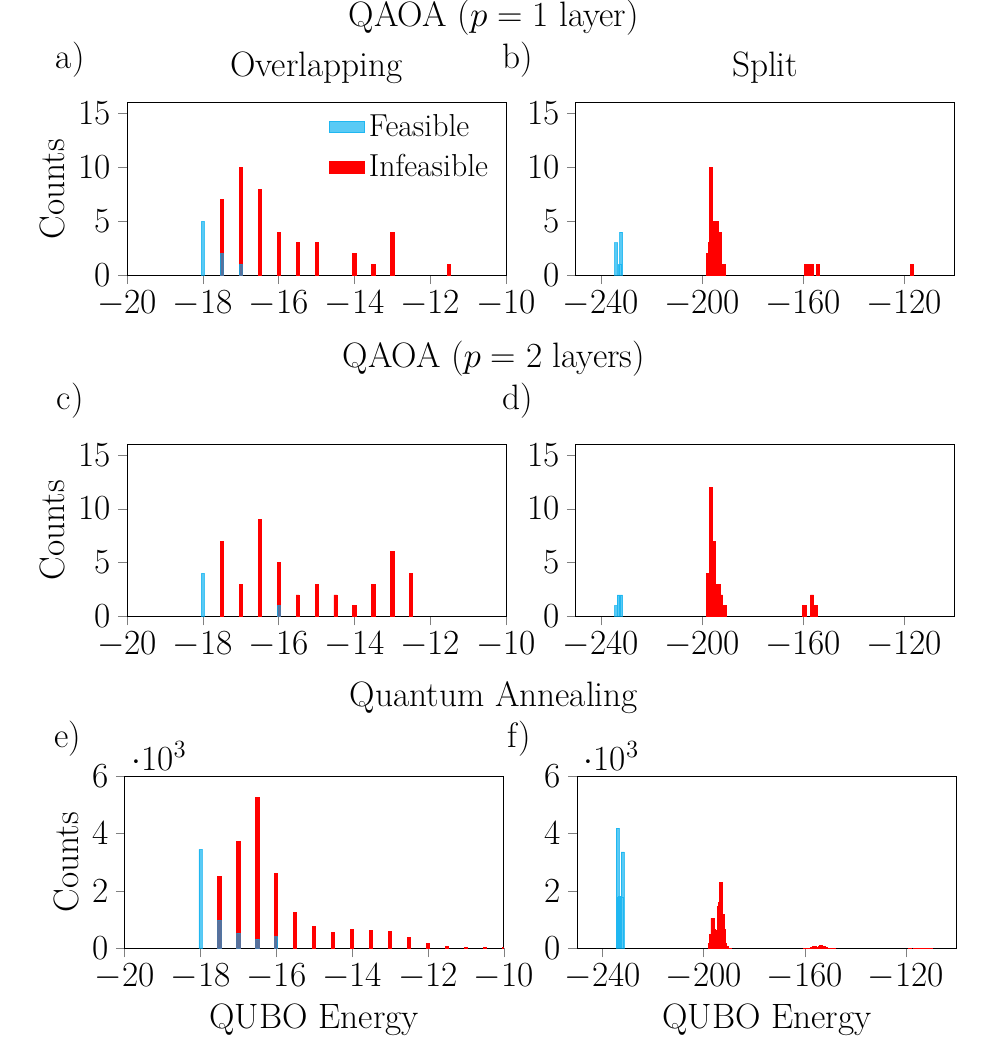}
	\caption{Histograms of the measured energies for one particular disturbed 18 variable problem with 2 trains from noisy simulations of the IonQ Aria-1 device running QAOA with a single-layer (a,b) or two-layer (c,d) ansatz and from experiments with the D-Wave annealer with a $10\mu$s annealing time (e,f). Results are shown for two choices of penalty values leading to overlapping (a,c,e) or split (b,d,f) spectra. Notice that on the Aria-1 device we obtain better results with a single-layer ansatz than with a two-layer ansatz, especially when the penalties lead to an overlapping spectrum.}
	\label{fig::energy_spectra}
\end{figure}

For one specific two train scheduling problem with a disturbance requiring 18 variables, a comparison of the results obtained with the IonQ Aria-1 device with both ansatze and the results obtained with the D-Wave annealer is presented in Fig.~\ref{fig::energy_spectra} (for an additional point of comparison, we present some results from simulations of IBM's superconducting devices in 
the Supplementary Information). 
In this case, it is clear that the single-layer QAOA ansatz performs better than the two-layer ansatz, especially when the feasible and infeasible spectra overlap. 
An interesting qualitative observation is that while there is some similarity in the shapes of the histograms of result energies obtained with the different platforms, they are nonetheless visibly different. This is true for the different variational ansatze on the Aria-1 device as well, especially in the distribution of feasible solutions. To explain the potentially better performance in the cases with overlapping spectra, recall that high penalty values can reshape the low energy part of the optimization landscape. This potentially introduces gaps before good feasible solutions, and hence may trap the quantum optimization in higher energy states -- see the Supplementary Information for further discussion including the entire spectra (split and overlapping) of selected QUBO problems.

\begin{figure}
	\centering
	\includegraphics[width = 1.0\columnwidth]{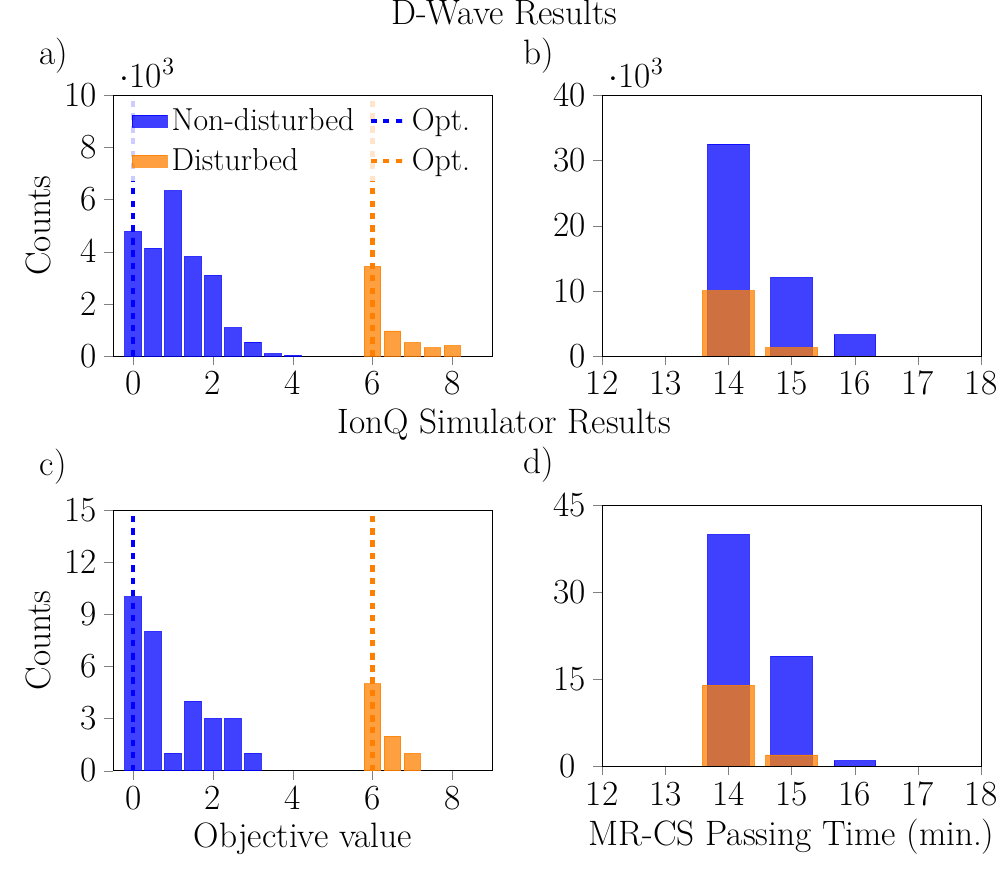}
	\caption{Histograms of the objective values and of the MR-CS passing times from schedules obtained from the D-Wave annealer with a $10\mu$s annealing time (a,b) and from noisy simulations of the IonQ Aria-1 device running QAOA with a single-layer ansatz (c,d). In both cases, the penalty values $p_{\text{sum}} = 4$ and $p_{\text{pair}} = 2$ were used, and only feasible solutions fulfilling all constraints from Section~\ref{sec::QUBO_model} have been counted. Both disturbed and non-disturbed problems are shown, with the optimal objective values indicated with dashed lines. Recall that the ILP versions of the problems were solved using CPLEX, returning optimal results (single for each instance) with an objective value of $0$ and $6$ (and $14$ minutes of MR-CS passing time) in computational time of $0.0016$ seconds.}
	\label{fig::qubo_small_DWave_gates}
\end{figure}

Fig.~\ref{fig::qubo_small_DWave_gates} continues the comparison between the results obtained with the D-Wave machine and the IonQ simulator running QAOA with the one-layer ansatz, now focusing on the properties of the solutions to the two train 18 variable railway scheduling problem, both with and without disturbances. Here, we show a comparison of the objective values (the tardiness) associated with the feasible solutions returned by each device, alongside the passing times from ${\rm MR}\rightarrow{\rm CS}$ encoded into each solution.
We observe that the distribution of passing times produced by the two approaches are very similar, despite the distributions of objective values being quite different.

\begin{figure}
	\centering
	\includegraphics[width = 1.0\columnwidth]{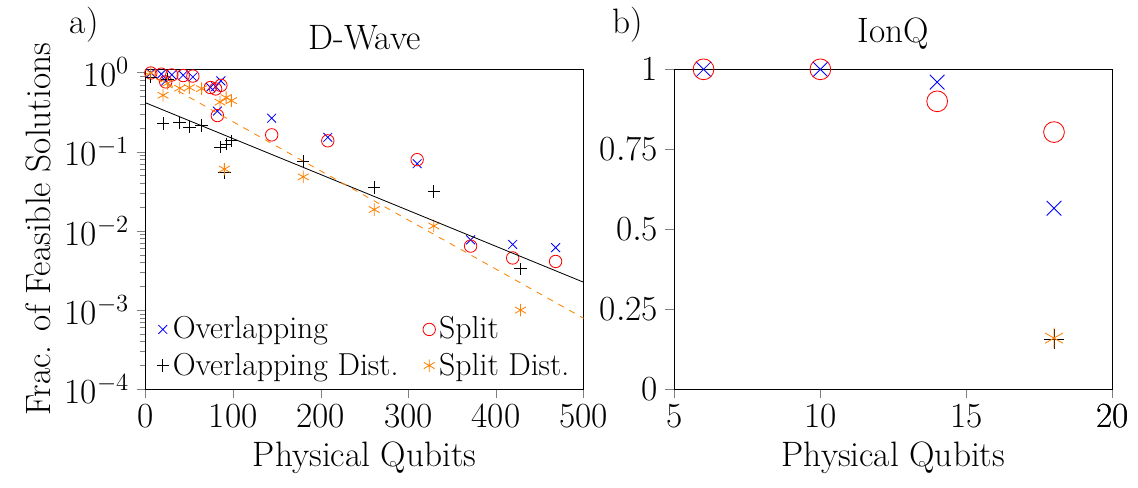}
	\caption{Evolution of the fraction of solutions to scheduling problems which are feasible as a function of the problem size, measured by the number of physical qubits required to embed the associated QUBOs on each platform. For the D-Wave annealer (a), the results are taken from experiments with an annealing time of $T=10\mu$s. Fits are shown for the disturbed problems with penalty values corresponding to both the overlapping and split cases, which shows that the fraction of feasible solutions decreases approximately exponentially with increasing problem size. For the IonQ Aria-1 device (b), the results are taken from noisy simulations of QAOA run with a single-layer ansatz. Note that there is no embedding overhead on Aria-1, each variable is directly associated with one physical qubit.
	}\label{fig::scaling}
\end{figure}

\subsection{Scaling}

An important consideration for any proposed application of NISQ devices is scaling behavior, as that will indicate the size of problem instances that can be handled at present and inform what problems are likely to be accessible in the near future.
To this end, in Fig.~\ref{fig::scaling} we plot the fraction of schedules returned as solutions to our railway scheduling problems which were feasible (i.e., those which fulfill all constraints listed in Sec.~\ref{sec::QUBO_model}) as a function of the number of physical qubits required for their solution.
With both IonQ's gate-based simulator and D-Wave's quantum annealer, we unsurprisingly observe that larger problems are more likely to yield purported solutions which violated one or more constraints. For the D-Wave device, we have sufficient statistics to infer that the fraction of feasible solutions decreases exponentially with the number of physical qubits necessary to embed the problem. Bear in mind that the translation of the scheduling problem to QUBO significantly increases the number of variables due to the time discretization (Eq.~\eqref{eq::bin_v}), which is not necessary in the original ILP formulation. To reflect this in practice, we have solved the deterministic ILP problem with CPLEX (Python API version: 22.1.1.0 on the local CPU). In our case, the CPLEX solution was always optimal and the CPLEX computational time was short ($0.0014$ s to $0.08$ s), pinpointing the better performance of the ILP approach see \url{https://github.com/iitis/quantum-stochastic-optimization-railways/blob/master/solutions/cplex_benchmarks.json} for particular CPLEX results. However, our goal was to asses the accessibility of various NISQ devices to model actual railway stochastic system rather than to compare runtimes in finding optimal solutions.

Naively extrapolating to the sizes necessary to solve problems of practical significance in railway scheduling (e.g., scheduling a crowded metro \cite{zhou2022joint} which is roughly $60\times$ larger than the largest problem solved in this work) indicates that such problems will not be accessible even to large quantum devices without some other improvements. Nevertheless, there exist approaches such as that of multilevel combinatorial optimization \cite{ushijima2021multilevel} based on hybrid computation which can allow the solution of problems an order of magnitude larger than current quantum devices.
To this end, we examine the statistical distribution of the results produced by our experiments with a goal of using quantum computing as one component of a hybrid algorithm for railway scheduling.

\subsection{Real-world railway rescheduling perspective}

\begin{figure}
	\centering
	\includegraphics[width = 1.0\columnwidth]{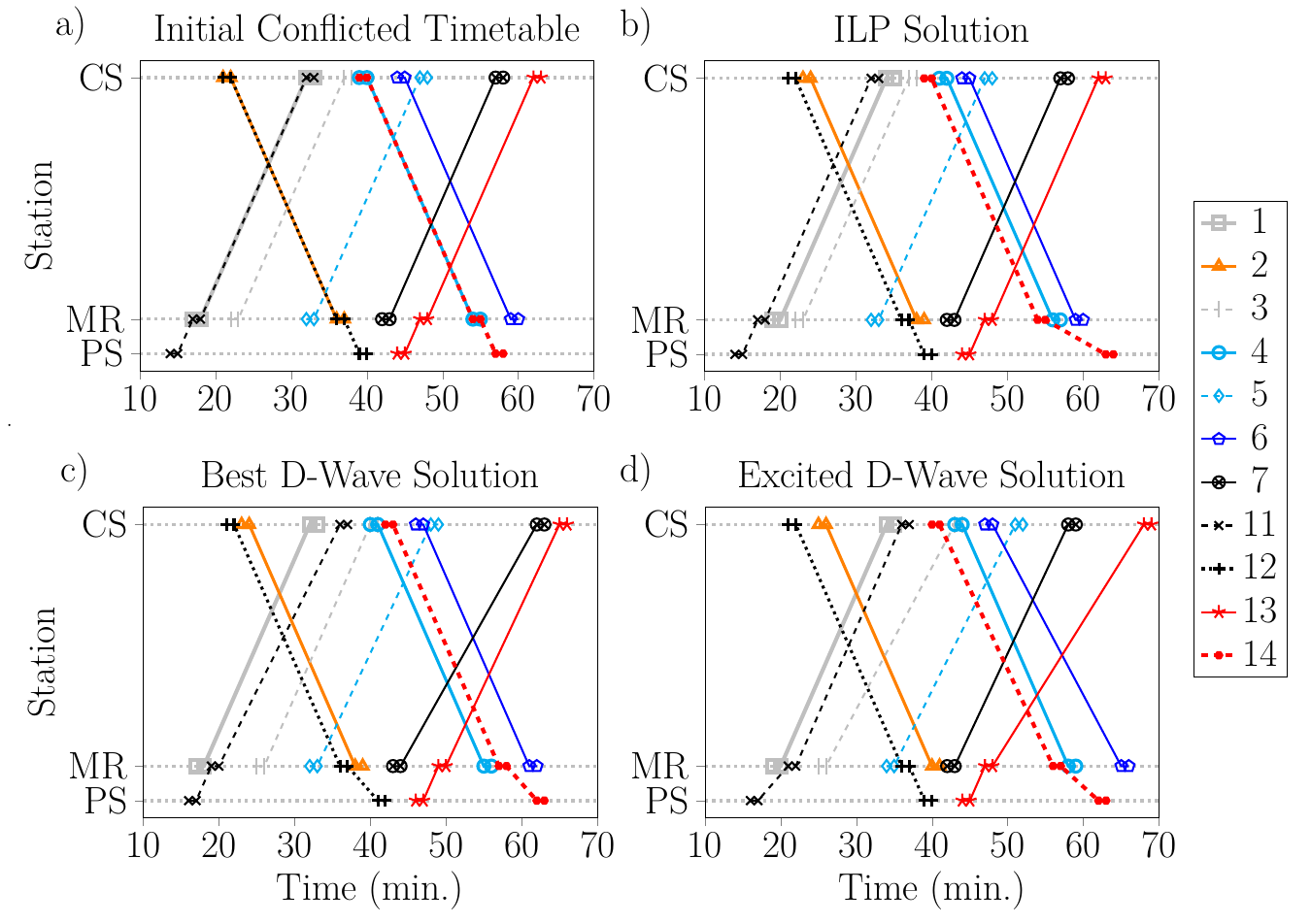}
	\caption{Train timetable diagrams illustrating an input disturbed timetable for an $11$ train schedule with conflicts (a) alongside several possible new solution timetables which resolve the conflict. Each line segment represents a train moving from station to station, and trains are numbered according to standard practice in railway scheduling. Shown is (a) an optimal solution found using ILP, (b) the best solution obtained with the D-Wave annealer (c), and one example of a highly-excited yet still feasible solution yielded by the annealer (d). For the D-Wave experiments, the annealing time was $10\mu$s and the penalty values were $p_{\text{sum}} = 4$, $p_{\text{pair}} = 2$ yielding an overlapping spectrum.
	}
	\label{fig::train_diagrams}
\end{figure}

To best connect our results with the practical perspective of railway scheduling and operation, we focus on the large $11$ train disturbed scheduling instance.
The associated $182$ variable QUBO can be solved on the D-Wave annealer (requiring $503$ physical qubits) but is too large for current gate-based quantum computers.

For this particular problem, the input disturbed timetable includes conflicts between three pairs of trains. A train diagram representing this timetable is shown in Fig.~\ref{fig::train_diagrams}a. 
To resolve this conflict, it is necessary that some of the trains involved in conflicts add an additional wait to their schedule. Fig.~\ref{fig::train_diagrams}b shows an optimal rescheduling that resolves these conflicts, generated with a classical ILP solver. Notice that passing times of all the trains are the same (i.e., the slopes of the lines are identical). This is not what would be expected for a timetable describing the movement of trains through stochastic areas of the rail network; in such a case, the passing times should be randomly perturbed and so the slopes should vary. This feature is visible in Fig.~\ref{fig::train_diagrams}c, which depict timetables corresponding to the best (lowest energy) solution returned by the D-Wave annealer and a solutions corresponding to a low-lying excited state, respectively. This observation supports the assertion that when NISQ devices are used to solve scheduling problems with this type of QUBO representation, their inherent noise may manifest in a way that can capture stochastic behavior of the underlying real-world model. (Finally, Fig.~\ref{fig::train_diagrams}d depicts one example of a highly-excited yet still feasible solution yielded by the annealer.)

\begin{figure}
	\centering
	\includegraphics[width = 1.0\columnwidth]{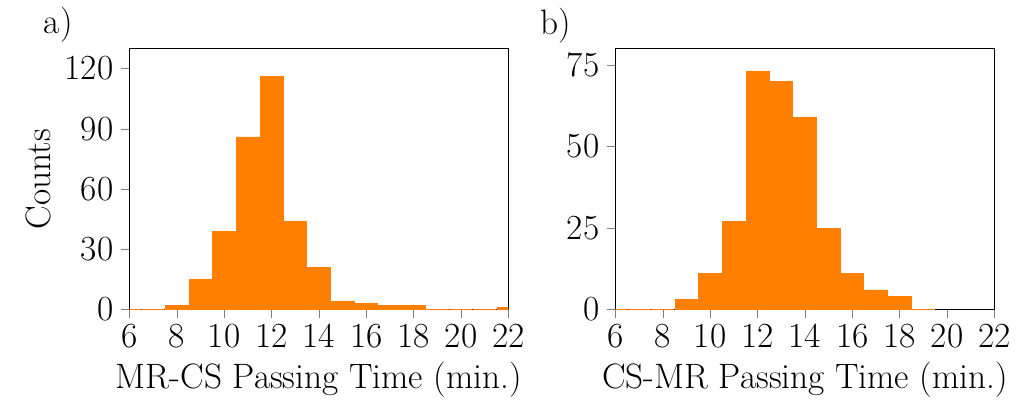}
	\caption{Histogram of actual passing times between Mount Royal (MR) and Camden Station (CS) for Baltimore Light RailLink trains in morning peak hours (7 a.m. - 10 a.m.) and afternoon peak hours (3 p.m. - 6 p.m.). Data for (a) northbound and (b) southbound trains are displayed separately. Data were collected for all workdays from 11 through 31 January 2024, excluding the 17th, 18th, and 26th due to Baltimore Light RailLink issues. 
	}
	\label{fig::real_data}
\end{figure}

To elaborate this observation, we compare the statistics of the D-Wave output to the statistics measured from real-world rail traffic. For this comparison, we present in Fig.~\ref{fig::real_data} the distribution of passing times northbound ${\rm MR}\rightarrow{\rm CS}$ and southbound ${\rm CS}\rightarrow{\rm MR}$. The data \cite{SwiftlyData}
was collected by aggregating passing times at peak hour during work days, as that is when the random disturbances due to road traffic are most intense. 

We observe that there is some qualitative similarity between the right tails of these distrubutions and the results from the D-Wave annealer as were shown in Fig.~\ref{fig::qubo_larger_dmax6}. The most common passing time in the real data is 12 minutes compared to 14 minutes in the D-Wave results, though this difference is due to specifics of Baltimore Light RailLink timetable and is also potentially influenced by the details of how the model was constructed and of how the real passing time is measured. In the real data we observe that some trains are recorded as going faster than the Baltimore Light RailLink timetable specifies. This is likely due to the inclusion of some extra reserve time in the Baltimore Light RailLink timetable. 

\begin{figure}
	\centering
	\includegraphics[width = 1.0\columnwidth]{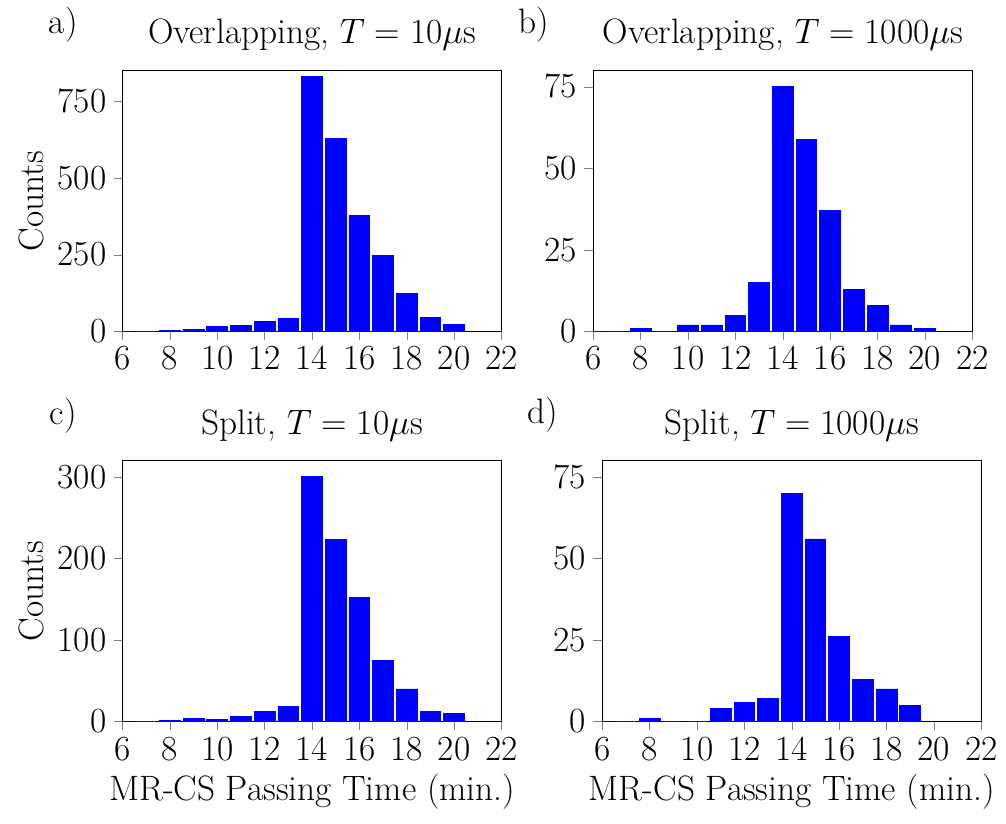}
	\caption{Histogram of MR-CS passing times obtained by solving the disturbed $11$ train problem with the D-Wave annealer. For this plot, the passing time constraint of Eq.~\eqref{eq::min_pass} was \emph{not} enforced when checking if a given solution was feasible. Results are shown for annealing times $T$ of $10\mu$s (a,c) or $1000\mu$s (b,d), and with penalty values set to $p_{\text{pair}} = 2$ and $p_{\text{pair}} = 4$ to produce an overlapping spectrum (a,b) or to $p_{\text{pair}} = 20$ and $p_{\text{pair}} = 40$ to produce a split spectrum (c,d).
		If we compare these distributions to those measured from actual rail traffic shown in Fig.~\ref{fig::real_data}, we see that while there are differences there are also qualitative similarities (e.g., in the shape of the right tail of the measured southbound distribution vs. the annealing results).}
	\label{fig::qubo_disturbed}
\end{figure}

To better reflect this behavior and so to capture the left tail of the real-world distributions, we alter our model by softening the constraints slightly. 
The transformation of the scheduling problem into QUBO form remains unaltered and the minimization proceeds as normal including all penalty terms, but at the end, we do not check any \textbf{passing time constraints} in Eq.~\eqref{eq::min_pass} when determining the feasibility of a solution. 
The distributions of passing times extracted from optimizations of the same $11$ train scheduling problem with the D-Wave machine following this relaxed proscription are shown in Fig.~\ref{fig::qubo_disturbed}. Note that the presence of solutions with passing times less than $14$ minutes from solutions which violate passing time constraints but respect the remaining constraints.
These solutions add a left tail to the distributions which brings them closer to the measured transit times shown in Fig.~\ref{fig::real_data}, especially with higher annealing times.
We expect that further tuning of the various parameters (especially the annealing time) can bring these distributions even closer together, and it is these similarities are the foundation for the modeling of stochastic rail traffic with NISQ devices.

As a final note, while we focus primarily on scheduling of train networks more specifically on scheduling the Baltimore Light RailLink network in this work, similar approaches can be followed for other types of transportation systems which will yield similar results. Distributions of typical passing times similar to those we present have been observed in road-based public transport, for example as shown in Fig.~1A of \cite{buchel2020review}. Moving forward, the techniques and approaches to rail scheduling discussed in this work to these and other types of public transportation networks and indeed to any analogous scheduling problems.

\subsection{Assessing costs of quantum computing}

To conclude our analysis, we assess the costs of the application of different paradigms of quantum computing for the real problem of Baltimore Light RailLink scheduling, both in terms of the compute time required to solve our scheduling problems and in terms of the monetary expense necessary to obtain access to the hardware. 

On the D-Wave machine, we ran each problem instance twice, once with $25000$ shots and an annealing time of $10\mu{\rm s}$ and once with the same number of shots and a longer annealing time of $1000\mu{\rm s}$. Combined, each problem instance consumed just over $25{\rm s}$ of compute time. Adding together the time taken to solve $30$ different scheduling problems -- and hence minimize $60$ different QUBOs (split and overlapping spectrum for each) -- in this way comes out to about $30$ minutes of time required, including overhead. 

As for our results from the IonQ Aria-1 device, the iterative optimization used in QAOA required the running of $\sim30-50$ circuits to evaluate the objective function for each QUBO solved. Unlike with the D-Wave machine, the time required for the circuits corresponding to the QAOA ansatz for different problems varies depending on the details of the scheduling problem, as the circuit depth varies. The runtime is dominated by two-qubit gates, which on the Aria-1 device take $600{\mu s}$ each. Most of the circuits run in the course of this work took on the order of $1-2{\rm s}$ to run 1024 shots, and each full optimization with QAOA required $40 - 60{\rm min}$. 

With IonQ's hardware, the expense of running a given circuit depends on the complexity of that circuit. For every optimization we ran but one, the circuits used were simple enough that they cost the minimum $97.50$ USD per run on the read device per circuit (with error mitigation enabled). The most complicated circuit (solving the 14 variable problem with the two-layer ansatz, 70 single-qubit and 126 two-qubit gates) cost $141.57$ USD per run.

We present a summary of the runtime requirements and approximate cost of the computations performed in this study in Tab.~\ref{tab::comp_costs}.
\begin{table}[ht]
		\centering
		\begin{tabular}{ccc}
			\hline\hline
			Quantum device& Total execution time  & Total cost (USD) \\ \hline\hline
			\makecell{D-Wave \\ Advantage\_system6.3} & $\sim\!30$\,minutes & $\approx \$1000$  \\
			\hline
			\makecell{IonQ  \\ Aria-1}  & $\sim\!21$\,hours & $\approx \$67000$ \\ 
			\hline\hline
		\end{tabular}
	\caption{Execution time and costs expended on each quantum platform for the computational results presented in this paper. The costs reported in the table are those incurred by us. The actual costs may differ for other users due to individual pricing policies of the companies.  Bear in mind, however, that a classical CPU can solve the ILP version of the problem to optimality more efficiently
		than the quantum devices considered and at significantly lower cost.}\label{tab::comp_costs}
\end{table}

\section{Conclusions}\label{sec::conclusions}

At present, quantum computing is in the era of NISQ devices, which are inherently noisy in an intrusive and unavoidable way. For this reason, it is important to incorporate some understanding of this noise at every level when attempting to build solutions for practical problems using these devices. Henceforth, leveraging quantum noise as a computational resource for optimization, rather than computational efficiency, was the primary goal of this work. In this work, we have demonstrated the use of current quantum computers in two different paradigms to study the dynamics of real-world rail or tramway operations. In other words, we have demonstrated the practical use of NISQ technology in solving real-world problems, particularly in optimizing transportation systems under stochastic disturbances. More specifically, we studied the problem of (re-)scheduling a set of trains which pass through portions of a rail network that have some degree of inherent stochasticity, e.g. due to sharing infrastructure with road traffic. By focusing on this inherently noisy optimization problem, we sought to actively exploit the noise present in all current hardware as a strength. Such research encourages collaboration across various fields, promotes innovation and new methodologies in both theoretical and applied research.

We focused on scheduling trains in a model of a portion of the Baltimore LightRail Link network, a tramway in Baltimore, Maryland, where between some stations trains must travel on tracks which share space with roadways leading to randomness in transit times.  Hereby, our research potentially improves the efficiency and reliability of public transportation systems, benefiting commuters and urban infrastructure. We constructed a variety of test scheduling problems starting from an actual timetable for the tramway, with the smallest problem having only a single train and the largest having twelve. These scheduling problems were easily translated to QUBOs, at which point we solved them using two types of NISQ devices. The first device employed was a quantum annealer, the D-Wave \texttt{Advantage\_system6.3}, which has 5614 qubits and could easily fit all of our scheduling problems. The second device was the IonQ Aria-1 gate-based trapped-ion computer, a 25-qubit device which was capable of solving several of the smaller scheduling problems we constructed using QAOA. While current gate-based computers have significantly fewer qubits than quantum annealers and while QAOA is a much more expensive approach to solving QUBOs than annealing, there is significant merit in the comparison. By testing our approach on two devices with completely different computational approaches, physical architectures, and noise models we can both demonstrate the broad applicability of these techniques as well as demonstrate how variations in the implementation alter the statistics of the output.

Our results show that when optimizing the schedules which govern trains passing through a stochastic zone, the effects of the random noise produce a set of schedules which when aggregated produce distributions of passing times which can potentially mimic the unpredictable delays visible in measurements of real railway traffic.
Consequently, this study can serve as a starting point for the development of specialized hybrid quantum/classical algorithms that integrate NISQ solvers alongside classical approaches to aid in capturing and optimizing the stochastic behavior of railway networks. This is an especially important point as large and practically-relevant railway scheduling problems \cite{zhou2022joint} are approximately $60\times$ larger than the largest problem we have successfully solved herein. By splitting the computation between a NISQ device which handles the noisy optimization component and a classical computer which handles the larger deterministic sections of the network which encompass and connect the stochastic zone(s), it will be possible to exploit the advantages of NISQ devices while avoiding limitations due to their small size. The development of effective error correction strategies, such as the one assessing the D-Wave embedding strategy, or error mitigation in variational quantum algorithms such as QAOA~\cite{botelho2022error} is a sound avenue for future work. Other future directions of work include the adoption of gate-based quantum optimization methods that propose better scalability in terms of the number of qubits \cite{tan2021qubit, tene2023variational} or that are are not based on QUBO formulations \cite{sciorilli2025towards}.

Finally, in demonstrating the potential for NISQ devices in the realm of stochastic scheduling we have taken an important step towards a resolution of several ongoing debates within the operational research community regarding the effectiveness of QUBO formulations such as the one we present in Sec.~\ref{sec::QUBO_model} for the planning and re-planning of railway schedules, and for more general practical applications~\cite{katzgraber2024searching} or similar problem formulations with more general QUBOs~\cite{glover2019quantum}. To this end, our research opens avenues for further research in quantum computing
applications, particularly in complex optimization scenarios and real-time system management.

\section*{Data availability}

The code and data used in this work have been made publicly available at \url{https://github.com/iitis/quantum-stochastic-optimization-railways}

Data from Baltimore Light RailLink has been collected on \url{https://github.com/iitis/Baltimore-Light-RailLink-data}

\section*{Acknowledgements}
B.G acknowledges support from the National Science Centre (NCN), Poland, under Project No.~2020/38/E/ST3/00269. 
K.D acknowledges: Scientific work co-financed from the state budget under the program of the Minister of Education
and Science, Poland (pl. Polska) under the name "Science for Society II" project number NdS-II/SP/0336/2024/01 funding amount 1000000 PLN  
total value of the project 1000000 PLN  \raisebox{-4pt}{\includegraphics[width = 0.15\textwidth]{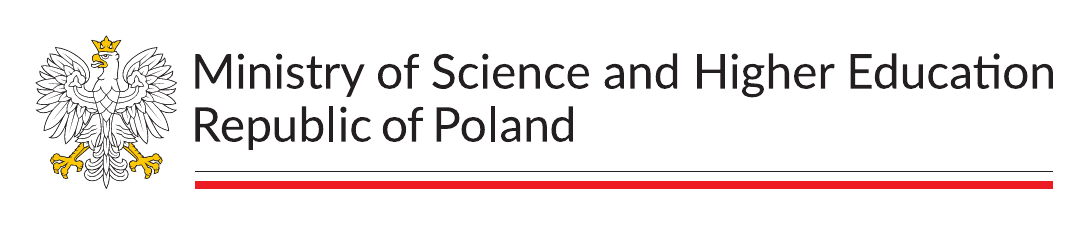}}.
S.D. acknowledges support the John Templeton Foundation under Grant No. 62422.
We acknowledge \emph{Swiftly's GTFS-realtime API} \url{https://swiftly.zendesk.com/hc/en-us} (accessed 11-31 January 2024) for supplying real-time traffic data.
K.D. acknowledges cooperation with Koleje Ślaskie sp. z o.o. (eng. Silesian Railways Ltd.) and appreciates the valuable and substantive discussions. 

\section*{Author contributions statement}

K.D., S.D. - conceptualization, K.D., E.D., R.R.  B.G. - preparing experiments, K.D., E.D., R.R. - running experiments,  K.D., E.D., R.R., B.G., S.D. - data analysis,  K.D., E.D. S.D. - manuscript writing, K.D., E.D., B.G., S.D. - manuscript supervision, S.D., B.G. - funding acquisition. All authors reviewed the manuscript.

\appendix

\appendix
\numberwithin{equation}{section}
\numberwithin{figure}{section}
\numberwithin{table}{section}

\section{Details of QUBO encoding}\label{app:qubo_encoding}

In this Appendix we present details on the encoding of the various constraints relevant for train scheduling in QUBO form, alongside an illustrative example demonstrating this encoding for a specific scheduling instance.

\subsection{Constraint encoding details}
QUBO constraints are derived directly from the Integer Linear Programming model in 
Section~II\,A, these are:
\begin{itemize}
\item The \textbf{minimal passing time} constraint: train $j$ may pass between subsequent stations $s \rightarrow s'$ no faster than $\delta^{pass}_{s,s'}$, and must respect the minimum station stay time $\delta^{\text{station}}$ 
[cf. Eq.~(7)]. 
For every train $j$ and every pair of stations $(s,s')\in SP_j$ this constraint takes the form:
\begin{equation}
    \sum_{t \in R_{s,j}} \left( \sum_{t' \in R_{s',j}, t' < t + \delta^{\text{station}} + \delta^{pass}_{s,s'} } x_{s,j,t} x_{s', j, t'} \right) 
    = 0.
\label{eq::bin_min_passing_time}
\end{equation}
\item The \textbf{minimal headway} constraint: two trains $j$ and $j'$ heading in the same direction must  be separated in time by $\delta^{\text{headway}}$, 
[cf. Eq.~(5)]. 
For all pairs of trains $(j,j')\in H_s$ which have this headway dependency when entering station $s$, the constraint reads:
\begin{equation}
  \sum_{ t \in R_{s,j}} \left(\sum_{ t' \in R_{s,j'}, t - \delta^{\text{headway}} < t' < t + \delta^{\text{headway}} } x_{s,j,t} x_{s,j't'} \right) 
  = 0.
  \label{eq::bin_min_headway}
\end{equation}
\item The \textbf{rolling stock circulation} constraint: for two trains $j$ and $j'$ with $j'$ following $j$ which share the same rolling stock, train $j'$ may not depart until train $j$ arrives, followed by the minimal preparation time $\delta^{\text{preparation}}$ and the minimal stay time $\delta^{\text{station}}$ 
[cf. Eq.~(6)]. 
For all pairs of trains $(j,j')\in RS_s$ that share rolling stock, the corresponding constraint is:
\begin{equation}
   \sum_{ t \in R_{s,j} } \left( \sum_{ t' \in R_{s,j'}, t' < t + \delta^{\text{preparation}} + \delta^{\text{station}} } x_{s,j,t} x_{s,j',t'} \right) 
   = 0.
   \label{eq::bin_rolling_stock}
\end{equation}
\end{itemize}

From the definition of the time range $R_{s,j}$ from 
Eq.~(2), 
we have that the outer sums in Eqs.~\eqref{eq::bin_min_passing_time}-\eqref{eq::bin_rolling_stock} include at most $d_{\max} + 1$ terms.
The number of terms in the inner sums varies from $1$ to $d_{\max} + 1$.
If we estimate that each inner sum has an average of $\sfrac{(d_{\max} + 1)}{2}$ terms, then we expect there to be an average of $\sfrac{(d_{\max} + 1)^2}{2}$ terms in total for each of the three constraint types given by Eqs.~\eqref{eq::bin_min_passing_time}-\eqref{eq::bin_rolling_stock}. 
From the discussions in 
Sec.~II\,A
concerning the size of the sets $SP_j$ and $H_s$, we expect the number of constraints to be linear in $\# J$ and $\# S$.
This statement relies on the reasonable assumption that there are few rolling stock constraints (i.e., that the sets $RS_s$ are small).

\subsection{Example instance: 2 trains, 18 variables}

Having described the implementation of the different constraint types, we now present details on the model of $2$ trains (corresponding to a disturbed schedule) with $d_{\max} = 2$ which requires 18 variables to encode.
The solutions to this instance as computed from D-Wave experiments and IonQ simulations are presented in 
Fig.~7 and Fig.~8.

For this example the non-disturbed timetable is as follows:
\begin{itemize}
\item Train $1$ is southbound, ${\rm PS} \rightarrow {\rm MR} \rightarrow {\rm CS}$, with timetable arrival times of: 
\begin{align}
    \tau^{\text{in}}_{1, {\rm PS}} &= 14 , \nonumber\\
    \tau^{\text{in}}_{1, {\rm MR}} &= 17 , \\ 
    \tau^{\text{in}}_{1, {\rm CS}} &= 32 . \nonumber
\end{align}
\item Train $2$ is northbound ${\rm CS} \rightarrow {\rm MR} \rightarrow {\rm PS}$, with: 
\begin{align}
    \tau^{\text{in}}_{2, {\rm CS}} &= 40 , \nonumber\\
    \tau^{\text{in}}_{2, {\rm MR}} &= 55 , \\ 
    \tau^{\text{in}}_{2, {\rm PS}} &= 58 . \nonumber
\end{align}
\end{itemize}
The disturbed input timetable we study here has train $1$ initially delayed by $5$ minutes, which places it in conflict with train $2$ through the \textbf{rolling stock circulation} constraint. 

\subsubsection{Building the QUBO}

Taking the maximal additional (secondary~\cite{DarianoPDPbb}) delay $d_{\max} = 2$, the ILP time variables from 
Eq.~(2)
for train $1$ are restricted to lie between:
\begin{equation}
\begin{alignedat}{2}
    l^{\text{in}}_{1, {\rm PS}} &= 19, \qquad\qquad& u^{\text{in}}_{1, {\rm PS}} &= 21, \\
    l^{\text{in}}_{1, {\rm MR}} &= 22, \qquad\qquad& u^{\text{in}}_{1, {\rm MR}} &= 24, \\
    l^{\text{in}}_{1, {\rm CS}} &= 37, \qquad\qquad& u^{\text{in}}_{1, {\rm CS}} &= 39, 
\end{alignedat}
\label{eq::qubo_ex_train1_times}
\end{equation}
and for train $2$ between:
\begin{equation}
\begin{alignedat}{2}
    l^{\text{in}}_{2, {\rm CS}} &= 40, \qquad\qquad& u^{\text{in}}_{2, {\rm CS}} &= 42, \\
    l^{\text{in}}_{2, {\rm MR}} &= 55, \qquad\qquad& u^{\text{in}}_{2, {\rm MR}} &= 57, \\
    l^{\text{in}}_{2, {\rm PS}} &= 58, \qquad\qquad& u^{\text{in}}_{2, {\rm PS}} &= 60. 
\end{alignedat}
\label{eq::qubo_ex_train2_times}
\end{equation}

From 
Eq.~(10)
there are $18$ total QUBO variables labeled $x_{s,j,t}$, where the subscripts run over the three stations $s\in\{{\rm PS},{\rm MR},{\rm CS}\}$, the two trains $j\in\{1,2\}$, and the three times possible for each variable. The ranges of the time subscripts are given by Eqs.~\eqref{eq::qubo_ex_train1_times} and \eqref{eq::qubo_ex_train2_times}, e.g., for the variables $x_{PS,1,t}$ the time index is $t\in\{19,20,21\}$. 

The requirement that each train leaves each station once corresponds to the one-hot constraints of 
Eq.~(14), written in terms of penalties in Eq.~(16). 
In our example, there are six such constraints which lead to six terms of the form:
\begin{equation}
p_{\text{sum}}  \Bigg( \sum_{\substack{t, t' \in R_{{\rm PS},1} \\ t \neq t'}} x_{\text{PS}, 1, t} x_{\text{PS}, 1, t'}  -\!\! \sum_{t \in R_{{\rm PS},1} } x^2_{\text{PS}, 1, t} \Bigg)
,
\label{eq::example_psum}
\end{equation}
yielding in total $54$ QUBO elements.

The \textbf{minimal passing time} constraint has the following form when written in terms of a penalty, shown for the case of train $1$ passing between stations ${\rm PS}\rightarrow{\rm MR}$:
\begin{equation}
p_{\text{pair}} \!\!
\sum_{t \in R_{{\rm PS},1}} \sum_{\substack{t' \in R_{{\rm MR},1}\\ t' < t + 1 + 2}} \!\!  \Big(x_{{\rm PS},1,t} x_{{\rm MR}, 1, t'} + x_{{\rm MR}, 1, t'} x_{{\rm PS},1,t} \Big)
,
\label{eq::example_mpt}
\end{equation}
there are four such equations (train $1$ passing ${\rm PS}\rightarrow{\rm MR}$ or ${\rm MR}\rightarrow{\rm CS}$, and train $2$ passing ${\rm CS}\rightarrow{\rm MR}$ or ${\rm MR}\rightarrow{\rm PS}$), which in total yield 24 QUBO elements.

The \textbf{rolling stock circulation} constraint written in terms of a penalty for the case of station ${\rm CS}$ takes the form:
\begin{equation}
p_{\text{pair}} \!\!
  \sum_{t \in R_{{\rm CS},1}} \sum_{\substack{t' \in R_{{\rm CS},2} \\ t' <  t + 3 + 1 }} \!\! \Big( x_{\text{CS}, 1, t} x_{\text{CS}, 2, t'} + x_{\text{CS}, 2, t'} x_{\text{CS}, 1, t} \Big),
  \label{eq::example_rsc}
\end{equation}
yielding $12$ QUBO elements. 

The objective function which is to be minimized is the tardiness:
\begin{equation}
\begin{split}
  f(\vec{x}) = &\sum_{ t \in R_{{\rm MR},1} } \frac{t - 17}{2} x_{{\rm MR}, 1, t} +
  \sum_{t \in R_{{\rm CS},1} } \frac{t - 32}{2} x_{{\rm CS}, 1, t}
  \\
   + &\sum_{ t \in R_{{\rm CS},2} } \frac{t - 40}{2} x_{{\rm CS}, 2, t} + \sum_{t \in R_{{\rm MR},2} } \frac{t - 55}{2} x_{{\rm MR}, 2, t} 
   ,
  \end{split}
  \label{eq::toy_objective}
\end{equation}
where the vector $\vec{x}$ is shorthand for all 18 variables $x_{s,j,t}$.
To translate this objective into a quadratic form suitable for a QUBO, we use the observation that $x_{s,j,t} = x^2_{s,j,t}$ since the variables are binary. 
The quadratic form $Q$ comes from combining the penalty-encoded constraints of Eqs.~\eqref{eq::example_psum}, \eqref{eq::example_mpt}, and \eqref{eq::example_rsc} with the objective function Eq.\eqref{eq::toy_objective}. In total there are 90 non-zero elements in this formulation of our two train scheduling problem.

\subsubsection{Solutions}

The example problem of this section is quite simple, and the resulting 18-variable QUBO is straightforward to solve with any classical solver. For the specific input data used for the disturbed input timetable where train $1$ is delayed such that there is a conflict due to \emph{rolling stock circulation} constraints, the optimal solution would be to let train $2$ wait one minute at station ${\rm CS}$, at which point the solution timetable may proceed with no conflicts. In terms of the $18$ variables used for the QUBO representation, the optimal solution has
\begin{equation}
\begin{alignedat}{3}
x_{{\rm PS}, 1, 19} &= 1, \quad& x_{{\rm MR}, 1, 22} &= 1, \quad& x_{{\rm CS}, 1, 37} &= 1,  \\
x_{{\rm CS}, 2, 41} &= 1, \quad& x_{{MR}, 2, 56} &= 1, \quad& x_{{\rm PS}, 2, 59} &= 1,
\end{alignedat}
\end{equation}
and all other variables equal to zero.
The value of the objective function from Eq.~\eqref{eq::toy_objective} is $f(\vec{x}) = 6$. This ground state is degenerate, as delaying train $2$ by one minute at $PS$ ($x_{PS, 2, 59} = 1 \iff x_{PS, 2, 60} = 1$) would yield the same value of the objective function and of the quadratic form.

An exhaustive enumeration of all $2^{18}$ possible values for the vector of variables $\vec{x}$ for this problem gives that the objective function may only take the values $f(\vec{x})\in\{6.0, 6.5, 7.0, 7.5, 8.0\}$ when considering only feasible solutions which do not violate any constraints. This is reflected in 
Fig.~8b. 
The solutions with $f(\vec{x}) = 6.0$ or $8.0$ are doubly-degenerate, as the delay at station ${\rm PS}$ is not counted in the objective.

Among these possible solutions, two example non-optimal solutions are nearly identical to the optimal solution but where one train required one additional minute to pass between ${\rm CS}$ and ${\rm MR}$.
The first of these solutions has
\begin{equation}
\begin{alignedat}{3}
x_{{\rm PS}, 1, 19} &= 1, \quad& x_{{\rm MR}, 1, 22} &= 1, \quad& x_{{\rm CS}, 1, 37} &= 1,  \\
x_{{\rm CS}, 2, 41} &= 1, \quad& x_{{\rm MR}, 2, 57} &= 1, \quad& x_{{\rm PS}, 2, 60} &= 1,
\end{alignedat}
\end{equation}
and all other variables zero, giving $f(\vec{x}) = 6.5$.
The second solution has non-zero variables
\begin{equation}
\begin{alignedat}{3}
x_{{\rm PS}, 1, 19} &= 1, \quad& x_{{\rm MR}, 1, 22} &= 1, \quad& x_{{\rm CS}, 1, 38} &= 1,  \\
x_{{\rm CS}, 2, 42} &= 1, \quad& x_{{\rm MR}, 2, 57} &= 1, \quad& x_{{\rm PS}, 1, 60} &= 1,
\end{alignedat}
\end{equation}
and gives $f(\vec{x}) = 7.5$. 

\subsubsection{Spectral Properties}
\begin{figure}
  \centering
  \includegraphics[width = 0.95\columnwidth]{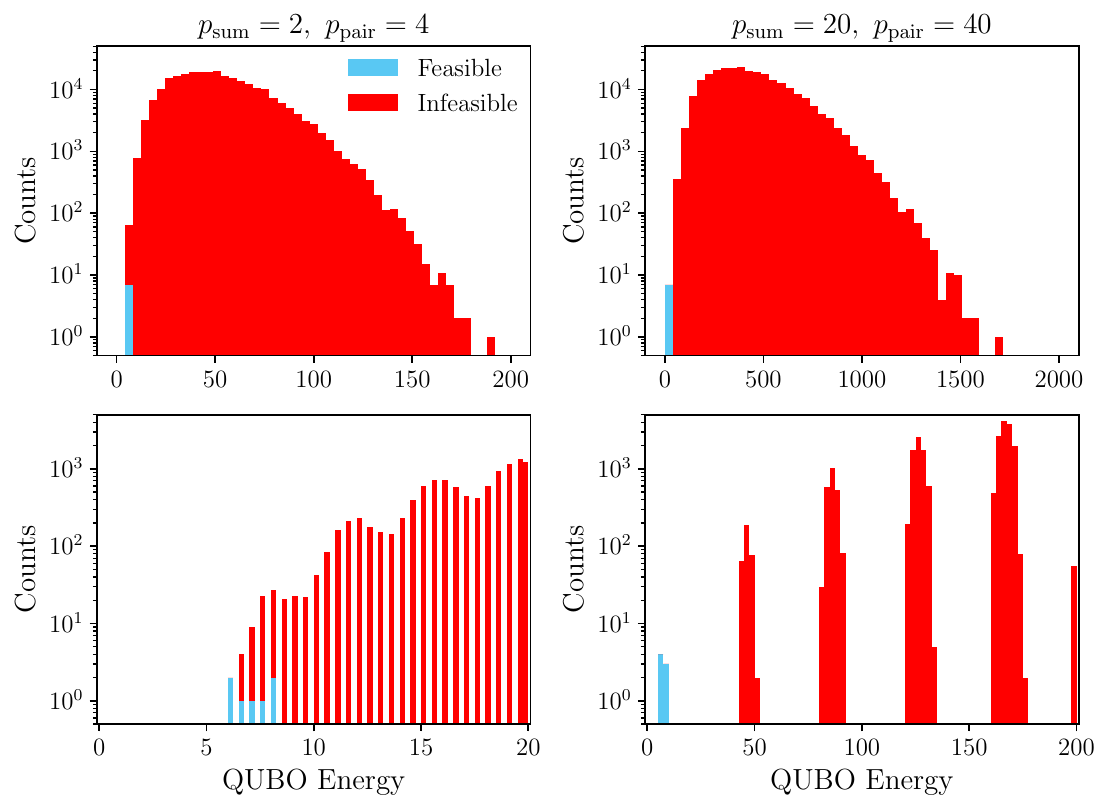}
  \caption{Histograms of QUBO values for the possible solutions to the 2 train, 18 variable problem with two choices of constraint parameters. The top panels include all $2^{18}$ possible solutions, and make it clear that there is little difference in the large-scale structure of the two choices beyond a rescaling of the energy. The bottom panels show only the feasible solutions as well as the infeasible solutions closest to optimal. With the lower choice of penalties, the two sets of solutions overlap, whereas with the larger penalties, the two sets of solutions are entirely separate.}
\label{fig::split_vs_unsplit}
\end{figure}

A complete and concrete specification of the QUBO to be minimized, formed from the combination of the constraints from Eqs.~\eqref{eq::example_psum}, \eqref{eq::example_mpt}, and \eqref{eq::example_rsc} alongside the objective from Eq.~\eqref{eq::toy_objective}, requires that specific values be chosen for $p_{\rm sum}$ and $p_{\rm pair}$. 
As discussed in Section 2.2, identifying good choices for these parameters can be challenging. 
In this work, we have chosen to consider each problem instance with two choices of penalty values, one ``overlapping'' case with lower penalties chosen such that the sets of feasible and infeasible solutions overlap in energy and one ``split'' case with higher penalties such that there is a moderate energy gap separating the two sets of solutions.
Figure~\ref{fig::split_vs_unsplit} shows the distributions of energies for these two choices applied to the 18-variable QUBO discussed in this section. 
On a broad, coarse-grained level the two distributions look essentially the same up to an overall rescaling of the energies, but the histogram of just a small window of energies around the optimal solution reveals the difference. 
With the smaller parameters, there is overlap between the sets of feasible and infeasible solutions and there are no large gaps (small gaps appear as all energies are integer multiples of $1/2$).
With larger penalty parameters, the distribution of energies reveals clusters of solutions with similar energies separated by large gaps. 
These clusters correspond to different constraint violations, hence all feasible solutions are isolated in their own cluster at the lowest values of the objective.

For the larger problems we consider it rapidly becomes intractable to exhaustively enumerate the collection of solution energies, but as is clear from the results obtained with the D-Wave annealer on an 11 train, 182 variable problem shown Fig. 3 this dichotomy between overlapping and split spectra are present in all the problem instances we consider.
Note that the same comb-like structure as appears in Fig. 3b and Fig. 3d would appear in the distribution of all energies in Fig.~\ref{fig::split_vs_unsplit} if the histograms used finer binning, as in the zoomed histograms.
When considering all energies a coarse binning is necessary for the feasible solution set to be visible -- the D-Wave annealer is by design biased toward sampling low-energy and feasible solutions.

\section{Hybrid algorithm for partially-stochastic railway networks}\label{app::hybrid_algorithm}

We propose a hybrid quantum-classical algorithm to handle railway or tramway re-scheduling problems that contain a stochastic component of limited size. In the proposed approach, the stochastic component will be optimized using a quantum device which will then be used as part of a classical optimization of the deterministic component. The final output will be a set of solutions derived from various possible outcomes from the stochastic component. Practically, this technique will be useful for a variety of real-world railway (re-)scheduling scenarios such as handling worst-case scenarios due to stochasticity, simulating the impact and spread of disturbances in the stochastic component, and providing decision-makers with a series of possible solutions to choose from.

As input, the algorithm takes a description of a railway scheduling problem. This consists of a statement of the topology, parameters, and constraints relevant to the network under consideration; an initial timetable which may include disturbances; an objective function to minimize; a decomposition of the network into stochastic and deterministic components, where the stochastic component includes a small portion of the network; and some details on the statistics of train traffic through the stochastic component of the network.

From this input data, the first step is to extract the sub-problem describing only the stochastic component. Following the approach discussed in this work, this sub-problem is transformed into a QUBO representation which can be solved on a quantum computer. Repeatedly solving this QUBO using a quantum computer or simulator produces a set of possible solutions, which can be filtered to yield a set of feasible solutions that satisfy all necessary constraints. For validation, the train traffic statistics can be computed from these solutions and compared to the specified distributions in the input data.

From this list of feasible solutions to the stochastic component some representative solutions are selected and combined with the remainder of the input data to construct a set of ILP representations of the scheduling problem on the deterministic component of the network, one per possible solution to the stochastic component. Each of these ILP problems is solved with standard classical techniques, and the best solution is determined. As it is possible that there is feedback between the stochastic and deterministic components of the network, this final solution is used in concert with the initial input data to construct a new stochastic scheduling problem. At this point, the algorithm is repeated in an iterative fashion, which may ultimately produce a joint solution with a better objective value. 

Finally, the output of the algorithm is a set of conflict-free solution timetables broken down by possible behavior in the stochastic component which covers both the stochastic and deterministic components of the network.

The proposed hybrid algorithm is presented below in bullet points:
\paragraph{Fixed inputs}
\begin{enumerate}
  \item Problem topology, parameters, constraints;
  \item Problem decomposition into stochastic (smaller) component and deterministic (larger) component;
  \item Train timetable, optionally including initial disturbances;
  \item Particular statistics of the train traffic on the stochastic part;
  \item Objective function;
\end{enumerate}
\paragraph{Processing}
\begin{enumerate}
  \item From initial data (timetable, disturbances), define the sub-problem on the stochastic component and encode it in QUBO form;
  \item Solve the QUBO on a quantum device;
  \item From QUBO solutions select feasible solutions, check the statistics of railway traffic parameters of the output, compare with expected statistics and re-sample or filter if necessary;
  \item Use selected QUBO solutions as inputs to the deterministic component encoded as an ILP problem, and run multiple optimizations with multiple inputs;
  \item Select a series of feasible solutions of the whole system;
  \item Prepare new QUBO for stochastic component given the solution of the stochastic component in point $5$, then repeat points and $2$-$5$ check whether objective improves;
  \item Execute point $6$ till stopping condition.
\end{enumerate}
\paragraph{Output}
\begin{enumerate}
  \item Series of solutions with various conflict-free timetables given various possible events in the stochastic component.
\end{enumerate}

\section{D-Wave quantum annealer details}\label{app:dwave_details}

\begin{table}
\centering
\begin{tabular}{|l|l|}
 \hline
\textbf{Parameter} & \textbf{Value} \\ 
 \hline
 \hline
Qubits & $5614$  \\  
\hline
 Couplers & $40105$   \\
 \hline
 Qubit temperature (mK) & $16.0 \pm 0.1$\\
 \hline
 Maximum achievable mutual inductance $\text{M}_{\text{AFM}}$ (pH)
 & $1.554$ \\
 \hline
 Quantum critical point (GHz) & $1.281$ \\
 \hline
 Qubit inductance $L_q$ (pH) & $382.180$ \\
 \hline
 Qubit capacitance $C_q$ (fF) &  $118.638$\\
 \hline
 Qubit critical current $I_c$ ($\mu A$) & $1.994$\\
 \hline
 Avg. single qubit thermal width (Ising units) & $0.221$ \\
 \hline
 FM problem freezeout (scaled time) & $0.073$ \\
 \hline
 Single qubit freezeout (scaled time) & $0.616$ \\
 \hline
 Initial value of external flux $\Phi^i_{\text{CCJJ}}$ ($\Phi_0$)
 & $-0.624$\\
 \hline
 Final value of external flux $\Phi^f_{\text{CCJJ}}$ ($\Phi_0$)
 & $-0.723$\\
 \hline
 Typical readout time, one qubit to full QPU  ($\mu$s)
 & $18.0 - 173.0$\\
 \hline
 Typical programming time ($\mu$s)
 & $\sim 14200$ 
 \\
 \hline
 QPU delay time per sample ($\mu$s) & $20.5$ \\
 \hline
 Readout error rate, full system & $\leq 0.001$ \\
 \hline
\end{tabular}
\caption{Physical properties of the D-Wave Advantage\_system6.3 machine \cite{dwave_machine}.}
\label{pegasus_prop}
\end{table}

For our experiments with quantum annealing, we made use of the D-Wave \texttt{Advantage\_system6.3} device. All the relevant physical parameters for this device are gathered in Table~\ref{pegasus_prop}. 

\begin{figure*}
    \centering
    \includegraphics[width= 0.45\textwidth]{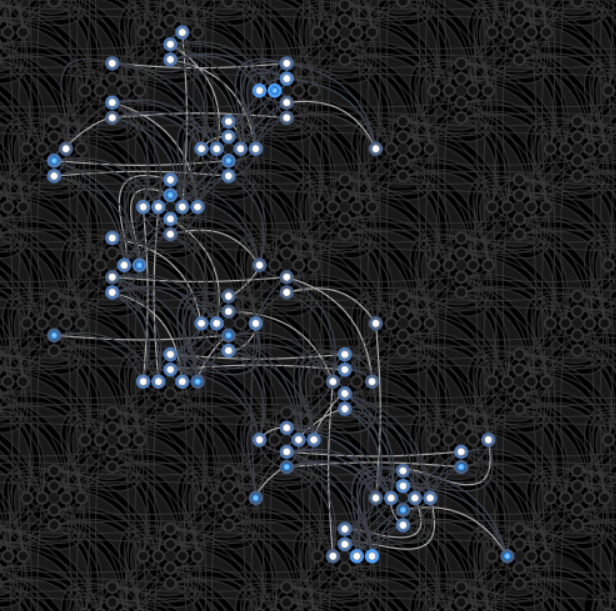}
    \hspace{1em}
    \includegraphics[width= 0.4585\textwidth]{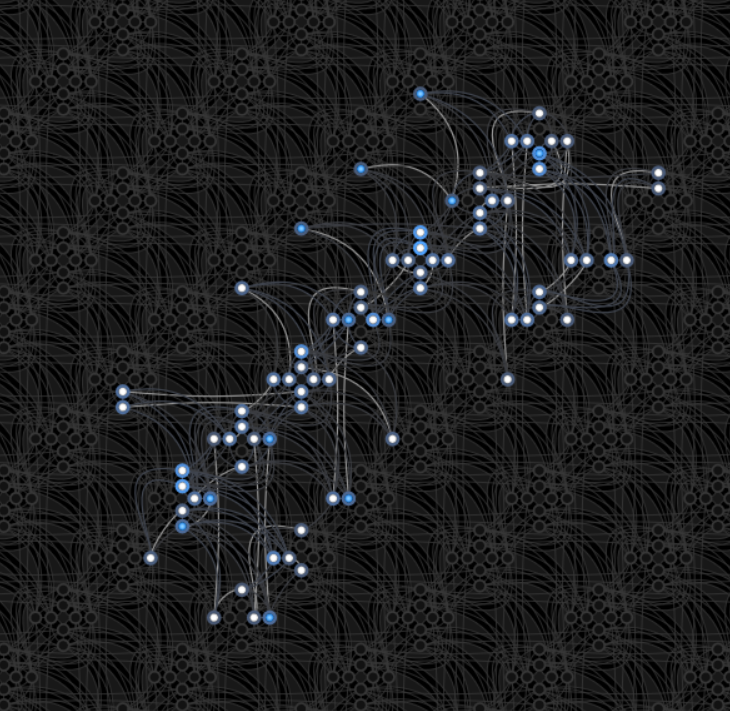}
    \caption{Illustration of the embedding of a two train scheduling problem with maximal secondary delay $d_{\max} = 6$, requiring $42$ variables into the native graph of the D-Wave annealer, for both disturbed (left) and non-disturbed (right) scenarios. After embedding, roughly 80 qubits were required.
    Dots represent active qubits and edges active coupling between qubits.}
    \label{fig::D-Wave_embedding}
\end{figure*}

This device contains 5614 superconducting qubits which are coupled according to the Pegasus graph structure \cite{dattani2019pegasus}. Since the native coupling graph of the annealer is quite far from being fully-dense, most problems must be modified to fit on the device through the use of an embedding procedure \cite{cai2014practical}. For example, one of the scheduling problems with $\#J=2$ trains and maximal secondary delay $d_{\rm max} = 6$ defined in 
Sec.~III
is mapped to a QUBO with 42 variables. After translating to an Ising model and embedding into the Pegasus graph, this problem required the use of approximately $80$ to $90$ qubits on the D-Wave machine (the embedding procedure is non-deterministic and small variations are expected). The final embedding for this problem is illustrated in Fig.~\ref{fig::D-Wave_embedding}.

The D-Wave embedding strategy uses chains of physical qubits to represent individual logical qubits. Here, we have used the default setting of \emph{dwave.system}, the \emph{EmbeddingComposite} Python library for all calculations on the D-Wave. The actual percentage of the number of solutions with no broken chains is greater than $50 \%$ for all calculations. In the worst case ($\# J = 11$ trains, maximal secondary delay $d_{\max} = 6$, and annealing time $10 \mu$s) we had still $55\%$ of solutions without chain break.

\section{IonQ quantum computer details}\label{app:ionq_details}

\begin{table}[h]
\centering
\begin{tabular}{ |l|l|l|}
 \hline
\textbf{Parameter} & \textbf{Aria-1} & \textbf{Brisbane}\\ 
 \hline
 \hline
Qubits & $25$ & $127$\\  
 \hline
$T_1$ & $100{\rm s}$ & $231.12\mu{\rm s}$ \\
 \hline
$T_2$ & $1{\rm s}$ & $145.97\mu{\rm s}$ \\
 \hline
Single-qubit gate time & $135\mu{\rm s}$ & \\
 \hline
Single-qubit gate error & 0.040\% & 0.025\% \\
 \hline
Two-qubit gate time & $600\mu{\rm s}$ & $660{\rm ns}$\\
 \hline
Two-qubit gate error & 4.250\% & 0.742\%\\
 \hline
Reset time & $20\mu{\rm s}$ & \\
 \hline
Readout time & $300\mu{\rm s}$ & $4\mu{\rm s}$\\
 \hline
SPAM accuracy & 99.520\% & \\
 \hline
 Readout error & & 1.320\% \\
 \hline
\end{tabular}
\caption{Characterization of the two gate-based quantum processors used in this work: IonQ's Aria-1 trapped-ion device \cite{IonqCloudConsole} and IBM's superconducting Brisbane device \cite{IbmQuantumDash}. The reported errors and lifetimes are average values for the Aria-1 device and median values for the Brisbane device.}
\label{ibmionq_prop}
\end{table}

As a representative of the gate-based approach to quantum computing, we chose to make use of the IonQ Aria-1 device. This is a gate-based quantum computer with 25 trapped Ytterbium ions acting as qubits, supporting all-to-all connectivity (meaning, two-qubit gates are directly supported between any pair of qubits). The various parameters which characterize the performance of the Aria-1 device are shown in Table~\ref{ibmionq_prop}. Note that the average gate errors are subject to significant variation over time, though they generally appear to be within a factor of $2\times$ in either direction of the numbers reported in the table. 

Finding the solution to a QUBO problem with the Aria-1 device with QAOA requires transforming the QUBO into a Ising model and finally into a circuit representing the QAOA ansatz which can be run on the device as part of a classical optimization loop. We used the Qiskit implementation of QAOA for this work, which transformed our input problems into circuits in terms of an abstract gate set which are then submitted to IonQ to be transpiled into the native Aria-1 gate set and run on the hardware.

\begin{figure*}
  \centering
  \includegraphics[width = 0.74\textwidth]{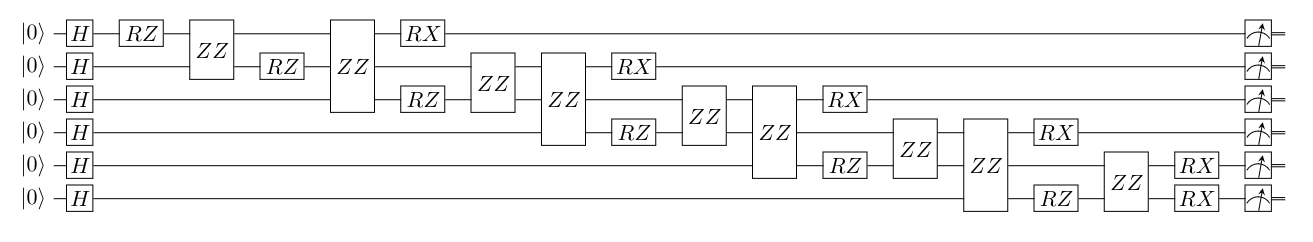}
  \caption{
  A quantum circuit diagram showing the single-layer QAOA ansatz for the simplest 6-variable QUBO, requiring 9 two-qubit gates (ZZ gates couple the outermost qubits the boxes cover) and a circuit depth of 17. This circuit is expressed in terms of an abstract gate set, and must be transpiled into the native gate set for any particular quantum processor for execution. The initial layer of Hadamard gates prepares the ground state of the initial Hamiltonian, then the layer of RZ and ZZ gates corresponds to the evolution under the cost Hamiltonian, and finally the layer of RX gates correspond to the mixer Hamiltonian. 
  }\label{fig::ionq1}
\end{figure*}

\begin{figure*}
  \centering
  \includegraphics[width=\textwidth]{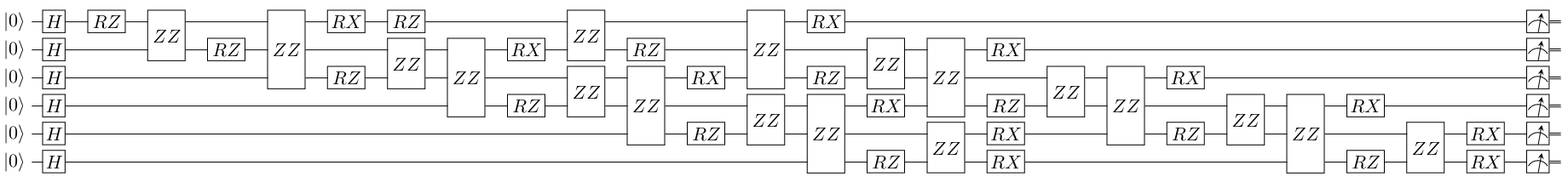}
  \caption{
  A quantum circuit diagram showing the two-layer QAOA ansatz for the simplest 6-variable QUBO, again in terms of an abstract gate set. This circuit contains 18 two-qubit gates and has a depth of 25. 
  }\label{fig::ionq2}
\end{figure*}

Unlike with the D-Wave machine, since the Aria-1 device has all-to-all connectivity there is no embedding overhead and thus the number of binary variables in the original QUBO is the number of qubits required.
For the smallest problem we considered, with one train and $d_{\rm max} = 2$, this number is six. The circuit implementing the single-layer QAOA ansatz for this problem is illustrated in Fig.~\ref{fig::ionq1}. In this circuit, the number of two-qubit gates (depicted as ZZ) required scales with the number of couplings in the Ising model (essentially, the number of off-diagonal non-zero matrix elements in the original $Q$ matrix defining the QUBO). The required circuit depth is not obvious from the figure, as we do not know the actual circuit run on the device after transpilation and so do not know if there are additional barriers or constraints to parallelism introduced at that stage. Nonetheless, we show in Fig.~\ref{fig::ionq2} the two-layer QAOA ansatz for the same problem in the abstract gate set. The number of two-qubit gates is exactly doubled since the circuit consists of two copies of the same subcircuit run one after another, however the figure shows that the required circuit depth could be significantly less than double. Since two-qubit gate errors are very likely the dominant error mechanism for the execution of these circuits this has little effect on the overall error, though it does indicate that the runtime need not double.

\section{Results with a superconducting IBM device}\label{app:IBM}

Another representative of the gate-based approach to quantum computing is provided by the superconducting quantum computers from IBM. Several of their 127-qubit \texttt{Eagle r3} devices are available through the IBM Quantum Platform, for example the Brisbane device whose properties are summarized in Table~\ref{ibmionq_prop}. 

In principle, solving our scheduling problems on these devices should follow essentially the same procedure as with solving them on the IonQ Aria-1 device: transform the QUBO into an Ising model, then build the QAOA ansatz from the Ising model in terms of the abstract gate set which is then compiled into the native gate set supported by the hardware. One source of complexity comes from the fact that on these devices, it is not possible to perform two-qubit gates between arbitrary qubits as the qubit connectivity graph is planar and only allows two-qubit gates between adjacent qubits. Effective gates between spatially-separated qubits must involve a series of qubit-qubit swaps until the gate can be performed between neighboring qubits. In this case, since the device has a highly-constrained connectivity graph it could be advantageous to follow some form of embedding procedure whereby one logical variable is spread over multiple qubits to allay connectivity issues, as is necessary on the D-Wave quantum annealer. We did not pursue this idea here, especially since the small scheduling problems examined are likely to not benefit.

Unfortunately, due to recent changes to how the IBM Quantum Platform operates and to the requirements which must be satisfied by circuits submitted to the cloud interface, the Qiskit implementation of QAOA we have used in this work can not currently be run on IBMs hardware \cite{IbmQaoaDoesNotWork}. Therefore, we can only present results from the noisy simulator which was provided as part of the platform until recently \cite{IbmNoMoreSim}. Using the \texttt{ibmq\_qasm\_simulator} configured to emulate the noise characteristics of the Brisbane device, we ran each of the three smallest scheduling problems (corresponding to 6, 10, and 14 variable QUBOs) five times, with two choices of penalty values yielding an overlapping or split spectrum. In terms of the abstract gate set, the circuits submitted were essentially the same as the ones used on the Aria-1 device, e.g. as in Fig.~\ref{fig::ionq1} for the 6 variable problem. Unlike with the Aria-1 device, however, it is likely that the transpilation process would add additional two-qubit gates over the minimum number set by the $Q$ matrix due to the restricted connectivity of the IBM device. For circuits with as few qubits as the ones we examine this is likely not a serious problem but for larger instances this overhead may dominate.

\begin{figure}
    \centering
    \includegraphics[width = 0.5\textwidth]{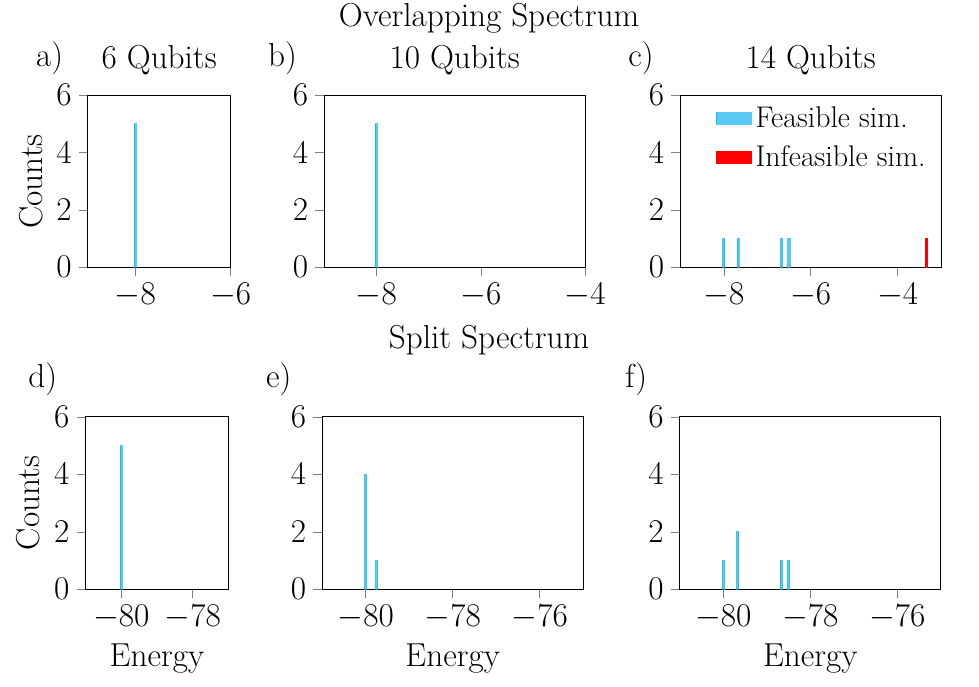}
    \caption{Histograms of the result energies yielded by noisy simulations of the IBM Brisbane superconducting gate-based quantum computer running QAOA with a single-layer ansatz. Each QUBO was minimized five times.
    }\label{fig::IBM_sim}
\end{figure}

The distributions of final energies returned from the noisy simulator are plotted in Fig.~\ref{fig::IBM_sim}. While it is impossible to draw reliable conclusions from so few data points, comparing these results to the equivalent results from the IonQ Aria-1 simulator seem to indicate that the performance of the IBM device should be expected to be comparable to the IonQ device, assuming that the simulator accurately captures the behavior of the device.

\end{document}